\documentclass[prb,twocolumn,showpacs,preprintnumbers,amsmath,amssymb,superscriptaddress]{revtex4}
\usepackage{graphicx}
\usepackage{dcolumn}
\usepackage{bm}
\usepackage{amsthm,amsfonts,graphicx,verbatim,color}
\usepackage{amssymb}
\usepackage{epstopdf}
\newcommand{\be}{\begin{equation}}
\newcommand{\ee}{\end{equation}}
\newcommand{\bea}{\begin{eqnarray}}

\newcommand{\eea}{\end{eqnarray}}

\newcommand{\der}{{\partial}}
\renewcommand{\phi}{\varphi}
\renewcommand{\epsilon}{\varepsilon}
\renewcommand{\vec}[1]{{\bf #1}}
\def\nn{\nonumber\\}

\begin{document}
\title{Orthogonal Metals:  The simplest non-Fermi liquids}
\author{Rahul Nandkishore}
\affiliation{Department of Physics, Massachusetts Institute of Technology, Cambridge, MA 02139}
\author{Max A. Metlitski}
\affiliation{Kavli Institute for Theoretical Physics, University of California, Santa Barbara, CA 93106}

\author{T. Senthil}
\affiliation{Department of Physics, Massachusetts Institute of Technology, Cambridge, MA 02139}  
\begin{abstract}
 
We present a fractionalized metallic phase which is indistinguishable from the Fermi liquid in conductivity and thermodynamics, but is sharply distinct in one electron properties, such as the electron spectral function. We dub this phase the `Orthogonal Metal.' The Orthogonal Metal and the transition to it from the Fermi liquid are naturally described using a slave particle representation wherein the electron is expressed as a product of a fermion and a slave Ising spin.  We emphasize that when the slave spins are disordered the result is not a Mott insulator (as erroneously assumed in the prior literature) but rather the Orthogonal Metal.  
We construct prototypical ground state wavefunctions for the Orthogonal Metal by modifying the Jastrow factor of Slater-Jastrow  wavefunctions  that describe ordinary Fermi liquids. We further demonstrate that  the transition from the Fermi liquid to the Orthogonal Metal can, in some circumstances, provide a simple example of a continuous destruction of a Fermi surface with a critical Fermi surface appearing right at the critical point. We present exactly soluble models that realize an Orthogonal Metal phase, and the phase transition to the Fermi liquid. These models thus provide valuable solvable examples  
 for phase transitions associated with the death of a Fermi surface.   
\end{abstract}
\maketitle
Despite tremendous effort in the last two decades our theoretical understanding of non-Fermi liquid phases of metallic matter in spatial dimension $d > 1$ is still in its infancy. Important examples of such non-Fermi liquids are provided by phases where the electron is fractionalized into multiple parts. Here, we present a new and particularly simple example of a fractionalized non-Fermi liquid phase, which behaves in every respect like a Fermi liquid, except that the charge carriers are orthogonal to the underlying electrons.  In some circumstances this phase is accessed via a direct second order phase transition from a Fermi liquid, which is marked by a sharp change in the electron spectral function. The transition  
 proceeds via a critical point that is characterized by a sharp critical Fermi surface where the Landau quasiparticle is critically destroyed\cite{Senthil}. 

To study this fractionalized non-Fermi liquid phase and the associated phase transition to a Fermi liquid, we employ a slave particle representation where the electron is expressed as a product of a slave Ising spin and a fermion. Unlike the more traditional `slave boson' representations, which have a U(1) gauge redundancy, and hence require introducing a compact U(1) gauge field \cite{Florens}, the slave spin representation has only a $Z_2$ gauge redundancy. A `slave spin representation' of this type was first introduced to study the multi-orbital Hubbard model \cite{Georges, Si, Ruegg, deMedici} and the possibility of orbital selective Mott transitions in such models. The slave spin formulation has gained in popularity over the last few years, and has been employed to describe correlation effects in multiband metals such as the iron pnictides, and also to investigate non-equilibirum physics in quantum quenches\cite{Fabrizio}.  However in the existing literature the phase transition where the slave Ising spin disorders has been 
 mis-identified as a  Mott (or in some cases orbital selective Mott) transition. Here, we point out that the correct identification is instead as a metallic phase in which all orbitals participate in transport and thermodynamics (just as in the usual Fermi liquid) but where some, possibly all, of the charge carriers are orthogonal to the physical electrons. For instance in the one band models considered in Ref. \onlinecite{Ruegg}  the zero temperature electrical conductivity is non-vanishing on both sides of the slave spin disordering transition, which should thus be interpreted as a transition between two metallic phases, one of which is a Fermi liquid, while the other is a fractionalized phase which has gapless charge carriers that are orthogonal to free electrons. The electron spectral function, in fact, has a gap even though the state is a compressible metal. We dub this phase an `Orthogonal Metal' to emphasize that the charge carrying fermions are orthogonal to the underlying electrons.  
 
 The Orthogonal Metal (and some simple generalizations described below)  is a minimal example of a non-Fermi liquid phase, quite possibly the simplest. Its low energy physical properties are readily determined reliably. The Orthogonal Metal is separated from the conventional Fermi liquid by a simple quantum phase transition driven by the condensation of the slave spin. In some situations this transition is second order, and the Landau quasiparticle of the Fermi liquid is critically  destroyed at the transition point. The properties of the resulting critical Fermi surface state can be straightforwardly obtained as we demonstrate later in the paper.

In the multi-orbital case (for instance Refs. \onlinecite{Georges,Si,deMedici})   the disordering of the slave spin of some orbital signals a fractionalization of the electrons corresponding to that orbital into a charged fermion and a gapped Ising variable. The resulting charged fermion continues to be gapless and contributes to thermodynamics and transport just as in a Fermi liquid. However the fractionalization of the electron leads to a loss of the Landau quasiparticle in the electron spectral function corresponding to that particular orbital.  We dub this phase an ``Orbital Selective Orthogonal Metal" and emphasize that it is {\em not} in an Orbital Selective Mott state. 

Two other generalizations are important to mention here. First there is no restriction on commensuration of the partice density with the underlying lattice for such a phase to exist. Thus we expect that the Orthogonal Metal phase can occur even for interacting electrons in the continuum. Clearly we can imagine generalizations of the simplest Orthogonal Metal to situations in which the slave spin is chosen to be some other discrete variable, for instance, a $Z_n$ rather than Ising, degree of freedom. In that case the gauge redundancy is also $Z_n$ and the corresponding Orthogonal Metal is obtained when the slave spins are disordered, the fermions form a gapless fermi surface, and the $Z_n$ gauge fields are gapped. For most of the paper we will
discuss only single band models and Ising slave spins as it most clearly and simply illustrates the main results. Generalization to other situations is straightforward and we will comment on these briefly when we discuss phase transitions to the conventional Fermi liquid.

 For further insight we show how the Orthogonal Metal may also be accessed within the more standard slave boson formulation\cite{Lee}. With this understanding we write down  prototypical wavefunctions for an Orthogonal Metal and show how they differ from the wavefunction of a conventional Landau Fermi liquid.
    
We also display exactly soluble models for the Orthogonal Metal phase and the associated transition to the Fermi liquid.  As a stepping stone to this end we first discuss a quantum Ising model on a square lattice which displays both a trivial paramagnetic phase and a non-trivial paramagnetic phase with $Z_2$ topological order. These phases are separated by a continuous confinement/deconfinement transition in the Ising universality class. Introducing electron degrees of freedom into this Ising model leads us to our desired exactly soluble model for an orthogonal metal phase and its phase transition to the conventional Fermi liquid.

We also consider a second exactly soluble model where the underlying quantum Ising model realizes both an Ising ferromagnetic phase and an Ising paramagnet which has non-trivial $Z_2$ topological order. This model has a soluble non-Landau transition between these two phases where the Ising ferromagnetic order parameter has critical correlations with a large anomalous exponent $\eta$. Such transitions have been studied for a number of years now\cite{LGWstar} though there have been no previous soluble examples that we are aware of. Introducing electron degrees of freedom, we again obtain an orthogonal metal phase and a ferromagnetically ordered Fermi liquid phase connected by a continuous transition. Thus we have constructed two soluble simple models for a phase transition associated with the continuous disappearance of the single particle Fermi surface. In both of these models, the transition between the Fermi liquid and the Orthogonal Metal proceeds via a critical Fermi surface.  A curious feature of the second model is that a second `mirror Fermi surface' appears at the critical point. We explain the origin of this mirror surface. We also point out that the critical Fermi surface of this second model encloses a volume that violates the Luttinger theorem (by a factor of two). This exactly soluble model thus also illustrates the fact that non-Fermi liquids with sharp Fermi surfaces need not satisfy the Luttinger theorem. 

Previous tractable examples of phase transitions where the electron Fermi surface is critically destroyed include the Kondo breakdown transition in Kondo-Heisenberg models \cite{Kondo1, Kondo2,paul}, and Mott metal-insulator transitions between a Fermi liquid and a spin liquid \cite{Mott,Mott3d}. In all these prior examples  the critical theory is characterized by the appearance of an emergent gapless $U(1)$ gauge field that couples to a Fermi surface and other gapless excitations. The example discussed in this paper 
provides a `baby' version of such a transition where there is an emergent gapped  gauge field which plays no direct role in determining universal critical properties. The properties of the resulting critical Fermi surface state can therefore be obtained rather straightforwardly. 

 The paper is structured as follows: first we provide a brief outline of the slave spin representation.  We demonstrate that in the low energy effective theory the electric charge is carried by the orthogonal fermions. We construct the current operator, and demonstrate that the ferromagnetic and paramagnetic phases of the slave spins 
are thus both metals, if the orthogonal fermions are in a metallic phase. We then discuss the relationship between the slave spin and the more usual slave boson formulations. This then leads to a discussion of prototypical wavefunctions  for the Orthogonal Metal phase. The Orthogonal Metal is argued to have a structure similar to a Slater-Jastrow wavefunction,  but with a Jastrow factor replaced by a different function. 

We then turn our attention to the phase transition, and explain that the transition between the two metallic phases is marked by a change in the electron spectral function. We show that in a mean field approximation, there appears a `critical Fermi surface' - a sharp Fermi surface that remains well defined, even when there is no Landau quasiparticle. The stability of the mean field description is determined by the relevance of the coupling of the energy density of the slave spins to fluctuations of the gapless modes of the  orthogonal fermion sectors at the critical point.
For continuum realizations of the Orthogonal Metal with Ising slave spins  in the presence of long range Coulomb interactions such coupling is irrelevant at the critical point, so that the mean field description is robust. However, for generic lattice systems or in the absence of long range Coulomb interactions the coupling is weakly relevant in two space dimensions, and marginally relevant in three, so that the mean field description necessarily fails close to the critical point, unless the bare coupling is fine tuned to zero, as is the case in the exactly soluble models presented in Sec. V. For a non-zero bare coupling, the OM-FL transition may be first order, and the existence of a critical Fermi surface cannot be guaranteed. Nonetheless, for generalized Orthogonal Metals with, say $Z_4$ slave spins, a continuous OM-FL transition with a robust mean field description and a sharply defined critical Fermi surface can be obtained for general lattice models. 

We then present exactly soluble models that realize a $Z_2$ fractionalized, `Orthogonal Metal' phase. Finally, we discuss the types of physical systems in which an Orthogonal Metal phase might be realized. 

\section{General Arguments}
\subsection{Introduction to slave spins}
\label{sec:slaveintro}
 In the slave spin representation, the electron operators are represented as a product of a fermion operator and a pseudospin operator. The works of Ref.\onlinecite{Georges, Si, Ruegg, deMedici} use two slightly different slave spin formulations. We will use the formulation of Ref.\onlinecite{Ruegg}  although our essential points apply equally to the formulation of Ref.\onlinecite{Georges, Si, deMedici}.  In line with Ref.\onlinecite{Ruegg}, we write the electron operator as %
\begin{equation}
c_{i \sigma} =  f_{i \sigma} \tau^x_{i} \label{eq: repn}
\end{equation}
where $c_{i\sigma}$ is the electron annihilation operator on site $i$, with spin $\sigma$, $f$ is a fermion operator, and $\tau^x$ is a Pauli matrix acting on an Ising pseudospin. 
Occupied states have $\tau^z = +1$ whereas the unoccupied/doubly occupied states have $\tau^z = -1$. To focus on the issues in the simplest possible context we consider a model with only one orbital - the extension to multi-orbital models is straightforward, and involves the introduction of a separate slave spin for each orbital. 

It may readily be seen that the slave spin representation expands the Hilbert space. The physical Hilbert space has two states per site per spin (the electron is either present or absent), whereas the slave spin representation has four states per site per spin (the fermion may be present or absent, and the slave spin may point up or down). To obtain a good representation of the original problem, we must restrict ourselves to the physical Hilbert space. This is accomplished by enforcing the constraint
\begin{equation}
\tau_i^z  = -(1 - 2 f^{\dag}_{\uparrow} f_{\uparrow})(1 - 2 f^{\dag}_{\downarrow} f_{\downarrow}) \label{eq: constraint1}.
\end{equation}
If the constraint Eq.\ref{eq: constraint1} is implemented exactly, then the slave spin representation is identical to the original fermion representation. In the low energy effective theory, however, the constraint is implemented on average, by the method of Lagrange multipliers. 

What types of phases and phase transitions can we access using the slave spin representation? The general approach is to start with an electron Hamiltonian, rewrite the electrons in terms of slave spins and $f$-fermions, and then decouple the spin and fermion sectors in a mean field approximation (implementing the constraint (\ref{eq: constraint1}) approximately, by the method of Lagrange multipliers). For example, we could start with the Hamiltonian
\begin{equation}
H = - \sum_{ij\sigma} t_{ij} c^{\dag}_{i\sigma} c_{j\sigma} + \sum_{ij} V_{ij} n_i n_j  - \mu \sum_i  n_i;  \label{eq: H}
\end{equation}
where 
$\mu$ is the chemical potential and $n_i = \sum_{\sigma} c^{\dag}_{i\sigma} c_{i\sigma}$ is the total electronic density on the site $i$. For some of our results it will be important to include the long range part of the Coulomb interaction in $V_{ij}$  but for now we will leave its detailed form unspecified.  
After writing this Hamiltonian in terms of the fermion and slave spin operators, using Eq.\ref{eq: repn}, we then decouple the $f$ fermions and the slave spins in a saddle point approximation, implementing the constraint Eq.\ref{eq: constraint1} by using Lagrange multipliers. One then obtains the coupled Hamiltonians 
\begin{eqnarray}
H_f &=& - \sum_{ij\sigma} t'_{ij} f^{\dag}_{i\sigma} f_{j\sigma} - \sum_{i\sigma} (\mu + 4 \lambda_{i}) f^{\dag}_{i\sigma} f_{i \sigma} \nonumber\\
&+& \sum_{ij} \big(V_{ij} +  2 \lambda_{i}\delta_{ij}\big) \big( \sum_{\sigma} f^{\dag}_{i\sigma} f_{i \sigma}\big)\big(\sum_{\sigma'} f^{\dag}_{j\sigma'} f_{j \sigma'}\big) \label{eq: Hf}; \\
H_I &=& - \sum_{ij} J_{ij} \tau^x_{i}\tau^{x}_{j}
 + \sum_{i} \lambda_{i} \tau^z_{i}; \label{eq: Hi}
\end{eqnarray}
We have defined renormalized parameters
\begin{equation}
 J_{ij} = \frac{1}{2} t_{ij} \sum_{\sigma} \langle f^{\dag}_{i\sigma}f_{j\sigma}\rangle + c.c.;  \qquad t'_{ij} = \langle \tau^x_{i} \tau^x_{j} \rangle t_{ij} \label{eq: teff}
\end{equation}
where $\langle...\rangle$ denotes the ground state expectation value and $\lambda$ is a Lagrange multiplier enforcing Eq.~\ref{eq: constraint1} on average. The coupled Hamiltonians in Eqs.~\ref{eq: Hf}, \ref{eq: Hi} must be diagonalized self consistently. 

Note that the Hamiltonian, after a mean field decoupling of the slave spins and $f$ fermions, takes the form of a standard fermion Hamiltonian coupled to a (generalized) transverse field Ising model. In the $f$ fermion sector, there can clearly be a large number of phases. As we will show, the $f$ fermions carry all the quantum numbers of the electron. In addition they carry a $Z_2$ gauge charge so that they cannot be directly identified with the physical electron. The Ising spin created by $\tau^x_i$ also carries $Z_2$ gauge charge but is electrically neutral and does not carry the physical electron spin. 

 In this paper, we consider situations where the $f$-fermions are in a Fermi liquid state.  When the Ising spin is also ordered the resulting phase is just an ordinary Fermi liquid of the underlying electrons. What if the Ising spin is disordered?  In previous literature, this phase has been mis-interpreted as a Mott insulator. In fact, we will demonstrate that it  corresponds to a $Z_2$ fractionalized metallic phase where the charge carriers are orthogonal to the underlying electrons - the Orthogonal Metal.  We note in passing that the Orthogonal Metal is just one of a large family of phases that may be constructed by $Z_2$ fractionalization of the electron. For example, in [\onlinecite{Ruegg2}] the $f$ electrons were placed in a quantum spin Hall state, which is insulating in the bulk but has gapless $f$ fermion edge states. When the slave spins are disordered, the edge states are orthogonal to the regular quantum spin Hall edge states \cite{Kane}. This phase could be dubbed an `orthogonal spin Hall state' or an `orthogonal helical metal' according to taste. 

\subsection{Interpretation of the slave spin results}
\label{sec:inter}
To correctly interpret the slave spin ordering transition, we must first establish whether the electric charge is carried by the $f$ fermions or the slave spins. This is because the slave spins and the $f$ fermions have independent dynamics, once the constraint Eq.\ref{eq: constraint1} is implemented only on average. It is therefore meaningful to ask whether it is the slave spins or the $f$ fermions that couple to the external electromagnetic field. 

The electric charge is the Noether charge that corresponds to a global U(1) phase rotation symmetry, while the electric current is the coupling of the system to a U(1) (electromagnetic) gauge field. The appropriate form of the current operator in the effective low energy theory must be determined based on symmetry, by determining how the slave spins and $f$ fermions couple to the U(1) gauge field. 

 In the slave spin representation Eq.\ref{eq: repn}, a U(1) rotation of the $c$ operator $c \rightarrow c \exp(-i\phi)$ can only be matched by a U(1) rotation of the $f$ fermion operator, since the operator $\tau^x$ is purely real. Accordingly, the effective low energy Hamiltonian for the $f$-fermions (Eq.\ref{eq: Hf}) exhibits a global U(1) symmetry under the transformation $f\rightarrow f \exp(-i\phi)$, but the effective low energy Hamiltonian for the slave spins (Eq.\ref{eq: Hi}) does not exhibit a U(1) symmetry. Since the electric charge is the Noether charge associated with a U(1) symmetry, it follows that the electric charge must be carried purely by the $f$ fermions, and cannot be assigned to the slave spins. From this it follows that any phase where the $f$ fermions form a  gapless Fermi surface must be conducting, since they carry the electric charge. 
 
We now illustrate this point by deriving the current operator in the slave spin representation. To obtain the form of the current operator, we promote the global $U(1)$ symmetry to a gauge symmetry, and determine how the theory couples to the gauge field. Since only the $f$ fermion Hamiltonian exhibits a U(1) symmetry, it follows that the current operator must be constructed exclusively out of $f$ fermions. If we assume that the $f$ fermion Hamiltonian takes the form (\ref{eq: Hf}), 
then a site dependent U(1) rotation $f_{i\sigma} \rightarrow f_{i\sigma} \exp(-i\phi_i)$ must be matched by a transformation $t'_{ij} \rightarrow t'_{ij} \exp(iA_{ij})$, where the gauge field transforms as $A_{ij} \rightarrow A_{ij} + \phi_j - \phi_i$. The electric current operator is the term in the Hamiltonian that couples to the U(1) gauge field $A_{ij}$. It must therefore take the form 
\begin{equation}
j_{ij} = i e \sum_{\sigma} (f^{\dag}_{i\sigma} t'_{ij} f_{j\sigma} - h.c.) \label{eq: current}
\end{equation}
The factor of $e$ converts the number current into the electrical current. This is of course the expected form of the current operator for a theory in which the $f$ fermions carry the electric charge.  
 If the charge carrying $f$ fermions are in a metallic phase, it follows inexorably that the state must have a non-vanishing electrical conductivity. This point is discussed further in the appendix. 
 
 Similarly it follows that the total particle number of the physical electrons is again given simply by the number of $f$-particles. These $f$-particles are coupled only weakly to the slave Ising spins. Thus a phase where the $f$-particles are compressible will also be compressible for the underlying electrons. We conclude that when the slave spins are disordered the result is a compressible metal rather than a Mott insulator. The Fermi surface of the $f$-fermions will be visible in quantum oscillation or other experiments that do not add/remove electrons. 
 
We now make a few comments regarding the low-energy effective theory of the Orthogonal Metal phase. The only degrees of freedom in this theory are the $f$-fermions in the vicinity of their Fermi surface. Thus, the low-energy theory for the $f$-fermions is precisely the same as Landau's Fermi liquid theory. Namely, the $f$-fermions interact via forward and BCS scattering interactions. Forward scattering interactions are exactly marginal and lead to renormalization of response functions such as e.g. compressibility and conductivity. BCS scattering is marginally irrelevant/relevant depending on the sign of the BCS amplitude. Attractive BCS interactions will lead to pairing of the $f$-fermions. The resulting state has been discussed in Ref.~\onlinecite{TSenthil}. Since the $f$-fermions carry physical electric charge $e$, the paired phase would be a superconductor with $hc/2e$ vortices. However, this state is distinct from a conventional superconductor as its ``Bogolioubov quasiparticles" are orthogonal to electrons and, in fact, carry an emergent $Z_2$ gauge charge. This phase, therefore, is topologically ordered. It was dubbed the $SC^*$ phase in Ref.~\onlinecite{TSenthil} to distinguish it from a conventional superconductor. 
 
 \subsection{Generalizations}
 \subsubsection{Multiband systems: Orbital selective Orthogonal Metals}
 Exactly the same issues as above arise for the slave spin descriptions of multi-orbital systems considered in Ref.\onlinecite{Georges, Si, deMedici}. Specifically these papers introduced a separate slave spin for each orbital and considered mean field states where the slave spins associated with some orbitals were disordered while other slave spins stay ordered. 
 Such states were interpreted to be examples of orbital selective Mott states where the electrons associated with some but not all orbitals in a multi-orbital system are localized. As the discussion above shows, in general, this interpretation is incorrect. Even when some of the slave spins disorder the $f$-bands will still contribute to thermodynamics and transport in exactly the same way as in the Fermi liquid metal. In particular quantum oscillation experiments will see a continuous evolution of the Fermi surfaces of all the bands across the slave spin disordering transition provided that the transition itself is continuous. (Here and below we assume that the Fermi surfaces of different bands do not intersect.) The only signature of the orbitally selective slave spin disordering transition is in photoemission experiments, and the disordered phase should be viewed as a phase with `orbitally selective orthogonality'.  \footnote{In some of the literature the generalized transverse field Ising models for the slave spins is solved in a further single site mean field approximation. This introduces some unnecessary artifacts such as making the bandwidth of the $f$-fermions zero.  For the generalized Ising model, this single site mean field approximation is strictly correct only in infinite dimensions; at any finite dimension there will be intersite correlations and the $f$-bandwidth will be non-zero.}
 
 \subsubsection{Orthogonal Metals in the continuum}
Mott insulating phases require the presence of an underlying crystalline lattice with which the electron density is commensurate. In contrast the Orthogonal Metal phase, being compressible, clearly has no such restriction on commensurability. Thus we expect that even in the absence of an underlying lattice ({\em i.e} with translation symmetry present) interacting electrons can display an Orthogonal Metal phase. As with the lattice realizations this phase will have low temperature thermodynamic properties identical to the Landau Fermi liquid but will have  a hard gap in the single particle spectral function. Quantum oscillation experiments will however see a conventional Fermi surface with volume determined by the usual Luttinger theorem. Moreover, for a Gallilean invariant system in the continuum, the parameters of the effective ``Fermi-liquid" theory of $f$-fermions will satisfy the standard relations, such as $m^* = (1+F^c_1) m$, with $m$ - the bare electron mass,  $m^*$ - the effective mass of the $f$-fermion, and $F^c_1$ - the forward scattering amplitude in the charge channel with angular momentum $\ell = 1$.

\subsubsection{Other discrete slave spins}
Clearly the construction above can be generalized to describe a family of Orthogonal Metals where the slave Ising spin is replaced by some other discrete spin variable. One example which we will consider below is when the slave spin is taken to be a $Z_n$ clock variable. This representation naturally leads to  generalized Orthogonal Metal phases where the electron fractionalizes into a charged spinful fermion and the $Z_n$ slave spin. Such generalized OM phases will be useful examples to consider when we discuss phase transitions between the conventional Fermi liquid and OM phases in Section \ref{ebmf} below. 

Unless otherwise specified we will restrict attention to the simplest OM phase in single band models with Ising slave spins in the rest of the paper.

 \section{Relationship to slave rotors}
It is useful to understand how the Orthogonal Metal phase may be described within  more standard slave particle frameworks where the electron operator is split into a charge $e$ spin-$0$ boson $b$ (the chargon) and a charge-$0$ spin $1/2$ fermion $d_{i\alpha}$ (the spinon):
\begin{equation}
c_{i\alpha} = b_i d_{i\alpha} \label{eq: slave bosons}
\end{equation}
An example is the slave rotor representation of Ref. \onlinecite{Florens}. As is well known this representation has a $U(1)$ gauge redundancy associated with letting 
$b_i \rightarrow b_i e^{i\theta_i}$, $d_{i\alpha} \rightarrow e^{-i\theta_i} d_{i\alpha}$ at each site $i$. Associated with this there is a local constraint 
\begin{equation}
n_{bi} = \sum_\alpha d^\dagger_{i\alpha} d_{i\alpha} \label{eq: constraint2}
\end{equation}
with $n_{bi}$ being the boson number on each site. 

If the $d_{i\alpha}$ form a Fermi surface and the $b_i$ condense then we get the usual Fermi liquid phase. At commensurate filling it is possible for the $b_i$ to form a Mott insulator while retaining the $d_{i\alpha}$ Fermi surface. This results in a description of an exotic quantum spin liquid Mott insulator which has a gapless spinon Fermi surface. 

Now consider a situation where the $b_i$ are not condensed but boson pairs are condensed, {\em i.e}, $\langle b_i \rangle = 0$ but $\langle b_i^2 \rangle \neq 0$. This reduces the $U(1)$ gauge structure to $Z_2$. The unpaired boson then survives as an Ising variable that carries the residual $Z_2$ gauge charge and is gapped in such a phase. However the boson pair condensate glues back the electrical charge to the spinon and converts it into a charge-$e$ spin-$1/2$ fermion that also carries $Z_2$ gauge charge. Thus we get back the slave spin representation of the electron operator. Since the single boson is not condensed the electron operator develops a gap in its spectrum. However the gapless Fermi surface now describes the $f$-particles and we get a compressible metal. This is just the Orthogonal Metal phase.

\section{Prototypical Wavefunctions}
The insights above enable us to write down prototypical wavefunctions for the Orthogonal Metal phase.  The many particle wavefunction for a fermion system must be antisymmetric under exchange. A class of such suitably antisymmetric wavefunctions can be easily constructed by multipliying the Slater determinant wavefunction of free fermions by a completely symmetric function of just the particle coordinates: 
\begin{equation}
\psi_{F}(\vec {r_1}\alpha_1, \vec{r_2} \alpha_2, .......\vec{r_{2N}}\alpha_{2N})  = \psi_b\left(\{\vec {r_i}\}\right)\psi_{Slater}(\{\vec{r_i}\alpha_i\}) \label{eq: electron wavefunction}
\end{equation}
Any choice of $\psi_b$ that is completely symmetric yields a legitimate electron wavefunction $\psi_F$. Being completely symmetric the function $\psi_b$ can be thought of as a wavefunction of a system of bosons. These bosons carry no spin (as $\psi_b$ does not depend on spin) and their coordinates are `slaved" to the coordinates of the fermions that form the Slater determinant. It is clear that there is a close connection between this class of wavefunctions and the slave boson description of interacting fermionic systems. 

The non-interacting Fermi gas is obtained if we choose $\psi_b$ to simply be the ground state wave-function of a non-interacting Bose condensate, 
\be \psi_b\left(\{\vec {r_i}\}\right) = 1 \label{BEC}\ee Correlations in a  Fermi liquid may be introduced by taking $\psi_b$ to be the wavefunction of an interacting bosonic superfluid.  A good choice is given by the Jastrow wavefunction
\begin{equation}
\psi_b\left(\{\vec {r_i}\}\right) \propto \prod_{i< j} f(\vec{r_i} - \vec{r_j})\label{eq:Jastrow}
\end{equation}
with $f(r) \to const$ as $r \to \infty$. This of course just yields the familiar Slater-Jastrow wavefunction for a Fermi liquid. 

Based on the discussion in the previous section we now see that a wavefunction for the Orthogonal Metal will be obtained if we choose the boson wavefunction to be that of a paired boson superfluid rather than an ordinary superfluid. Thus we write 
\bea
&&\psi_{OM}(\vec {r_1}\alpha_1, \vec{r_2} \alpha_2, .......\vec{r_{2N}}\alpha_{2N}) \nn
&&~~~~~~~~=\psi_{PSF}\left(\{\vec {r_i}\}\right)\psi_{Slater}(\{\vec{r_i}\alpha_i\})
\eea
where the paired boson superfluid wavefunction takes the form 
\begin{equation}
\psi_{PSF} \propto {\cal S} \left[g(\vec{r_1} - \vec{r_{2}})g(\vec{r_3}- \vec{r_{4}}).....g(\vec{r_{N-1}} - \vec{r_{N}})\right] \label{eq: pairing}
\end{equation}
Here ${\cal S}$ denotes the symmetrization operator, and $g(r_1 - r_2)$ is the wavefunction of a ``molecule" of two bosons satisfying $g(r) \to 0$ as $r \to \infty$.  Eq.~(\ref{eq: pairing}) represents a condensate of these molecules analogous to the single boson condensate in Eq.~(\ref{BEC}). Further correlations between molecules can be built in by a molecular Jastrow factor as in Eq.~(\ref{eq:Jastrow}).
 
\section{Electron spectral function: evolution through transition}
\label{esf}

We now demonstrate that the slave spin ordering transition is associated with a dramatic change in the nature of the electron spectral function (which may be measured directly through photoemission experiments). In particular, the ordered phase of the slave spins is a Fermi liquid, whereas the paramagnetic phase is an exotic metal where the charge carriers are orthogonal to free electrons.  Thus even though the thermodynamics and transport are identical to a Fermi liquid the disordered phase of the slave spins is not really in a Fermi liquid phase. Rather the electron spectral function acquires a hard gap everywhere in the Brillouin zone. (In the multi-orbital case the hard gap will appear just for the electron operator associated with the orbital whose slave spin was gapped).  We therefore dub this phase an Orthogonal Metal. The Orthogonal Metal is quite possibly the simplest non-Fermi liquid phase. Nevertheless it starkly illustrates how photoemission and quantum oscillation experiments may give seemingly completely contradictory results for the Fermi surface in a non-Fermi liquid metal. In the Orthogonal Metal, quantum oscillations will see a full Fermi surface while photoemission sees a hard gap. 

\subsection{Spectral function in the Orthogonal Metal}
In the present section, we describe the electron spectral function in the Orthogonal Metal using a mean field approximation that neglects any coupling between the slave spin and orthogonal fermion sectors. We expect this approximation to be at least qualitatively correct deep inside the OM phase.

To calculate the spectral function, we note that in Eq.\ref{eq: repn}, the electron operator is a product 
of a slave spin operator and a fermion operator. Therefore, in the mean-field approximation, the imaginary time electron Green function is the product of the slave spin Green function and the fermion Green function. In momentum space, the imaginary time electron Green function is then given by  the convolution of the Green functions for the slave spins and the $f$-fermions.
\begin{equation}
G(\vec{q}, i \omega) =  \int_\vec{k} \int_{\Omega} G_{spin}(\vec{q} - \vec{k},i\omega - i\Omega) G_{fermion}(\vec{k}, i \Omega) \label{eq: G}
\end{equation}
Here $G_{spin}$ and $G_{fermion}$ are the Green functions for the slave spins and $f$ fermions respectively. The electron spectral function $A$, defined as the imaginary part of the retarded Green function $G^R$,
\be A(\vec{q}, \Omega) = -\frac{1}{\pi} \Im G^R(\vec{q}, \Omega)\ee
may then be obtained by the analytic continuation,
\be G^R(\vec{q}, \Omega) = -G(\vec{q}, i \omega = \Omega + i \delta)\ee
Alternatively, the spectral function may also be found using the expression 
\begin{equation}
A(\vec{q}, \omega) = \int_\vec{k} \int_{0}^{\omega} d\Omega A_{spin}(\vec{q} - \vec{k},\omega - \Omega) A_{fermion}(\vec{k}, \Omega) \label{eq: spectral function2}
\end{equation}
Here $A_{spin}$ and $A_{fermion}$ are the spectral functions of the slave spins and the $f$ fermions respectively. Eq.\ref{eq: spectral function2} may be obtained from Eq.\ref{eq: G} by writing the imaginary time Green functions of slave spins and $f$ fermions in the spectral representation 
\begin{equation}
G(\vec{q},i\omega) = \int d\Omega \frac{A(\vec{q},\Omega) }{\Omega - i\omega} \label{eq: spectral representation}
\end{equation}
We may use either Eq.\ref{eq: G} or Eq.\ref{eq: spectral function2} according to convenience. 

\label{sec: spectral function}

In the mean field approximation and ignoring the effects of the interaction $V_{ij}$, the $f$ fermions are described by a free fermion hopping model, and therefore have the standard fermion Green function 
\begin{equation}
G_{fermion}(\vec{q}, i \omega) = \frac{1}{-i\omega + E(\vec{q})} \label{eq: fermion gf}
\end{equation}
where energies are measured relative to the Fermi energy.

The fermion Green function does not change across the slave spin ordering transition. However, the slave spin correlation function does change. In the ordered phase of the slave spins, the correlation function is a delta function in momentum space $G_{spin}(\vec{q}, i \omega) \sim m^2 \delta(\omega)\delta(\vec{q})$, where $m = \langle \tau^x \rangle$ is the mean magnetization of the slave spins. Convolving this with the fermion Green function Eq.\ref{eq: fermion gf}, continuing to real frequency and taking the imaginary part, we find that the electron spectral function in the ordered phase of the slave spins takes the form 
\begin{equation}
A(\vec{q}, \omega) = \langle \tau^x\rangle^2 \delta(\omega - E(\vec{q})).
\end{equation}
This spectral function is characteristic of a Fermi liquid, with the identification $Z = \langle \tau^x\rangle^2$, where $Z$ is the quasiparticle residue.

Meanwhile, in the paramagnetic phase of the slave spins the slave spin sector has a finite gap $\Delta$. Therefore, $A_{spin}(\vec{q}, \omega) = 0$ for $|\omega| < \Delta$ and so from Eq.~\ref{eq: spectral function2} the electron spectral function also satisfies $A(\vec{q}, \omega) = 0$ for $|\omega| < \Delta$. 
%
Hence, the electron spectral function has a gap equal to $\Delta$, in sharp contrast to the Fermi liquid. This phase is the Orthogonal Metal.

We now review the properties of the Orthogonal Metal phase. The Orthogonal Metal is a compressible and conducting phase of matter, which nonetheless has a spectral gap. In quantum oscillations experiments, a Fermi surface will be seen (the Fermi surface of the $f$ fermions). The thermodynamics are indistinguishable from those of a Fermi liquid. However, in photoemission, a single particle spectral gap will be observed, and no Fermi surface will be seen. 

The Orthogonal Metal is sharply distinct from the Fermi liquid only in its single particle properties. It follows that the transition from a Fermi liquid to an Orthogonal Metal will manifest itself sharply only in experiments that probe single particle physics, such as photoemission. We therefore study the evolution of the electron spectral function across the FL-OM transition. 

\subsection{Phase transition: Mean field theory}
\label{sec:AMF}
In this section, we describe the phase transition from the Fermi liquid to the Orthogonal Metal using a mean field approximation that neglects any coupling between the slave spin and orthogonal fermion sectors. In the following section we discuss under what circumstances the mean field approach provides an accurate description of the critical point.

 In the mean field approximation, the transition, approached from the Orthogonal Metal side, is marked by the closing of the spectral gap along a critical Fermi surface. Approached from the Fermi liquid side, the transition is marked by the vanishing of the quasiparticle residue. The slave spin ordering transition in mean field theory is thus a particularly simple example of a transition between metallic phases that proceeds via a critical Fermi surface \cite{Senthil}. 

Recall that in the Fermi liquid phase, the electron spectral function is 
\begin{equation}
A(\vec{q}, \omega) = Z \delta(\omega - E(\vec{q})),
\end{equation}
with $Z = \langle \tau^x\rangle^2.$
As we approach the transition by tuning a parameter (for instance, the interaction strength) $g$ towards a critical value $g = g_c$, the magnetization vanishes as $(g_c- g)^{\beta}$, where $\beta$ is the Ising model critical exponent\cite{Altland}. Thus, as we tune the interaction strength towards the transition, starting in the Fermi liquid, the quasiparticle spectral weight vanishes as 
\begin{equation}
Z \sim (g_c - g)^{2\beta}
\end{equation}
The vanishing of the quasiparticle residue at the slave spin ordering transition was first identified by Ref.\onlinecite{Georges}.

Meanwhile, in the Orthogonal Metal phase, the spectral function has a hard gap $\Delta$. As we tune towards the critical point, the gap closes everywhere on a momentum space surface that corresponds exactly to the $f$-fermion Fermi surface (and also to the non-interacting electron Fermi surface). Thus, the  electron Fermi surface is a privileged surface even when the transition is approached from the paramagnetic phase of the slave spins - it is the surface in momentum space at which the spectral gap closes. Note that since the electron gap $\Delta$ is the same as the slave spin gap, near the transition
\be \Delta \sim (g - g_c)^\nu \ee
with $\nu$ - the correlation length exponent of the Ising model.

At criticality, the slave spin spectral function takes the form
\begin{equation}
A_{spin}(\vec{q},\omega) \propto \frac{\theta(\omega^2 - c^2 q^2) \mathrm{sgn}(\omega)}{(\omega^2 - c^2q^2)^{1-\eta/2}}\end{equation}
where $\eta$ is the anomalous exponent of the Ising model. Substituting into Eq.\ref{eq: spectral function2} we find
\begin{equation}
A(\vec{q}, \omega) \sim \int_\vec{k} \frac{\theta\big((\omega - E_\vec{k})^2 - c^2 (\vec{q}-\vec{k})^2\big) \theta\big(E_\vec{k}\big) \theta\big(\omega - E_\vec{k}\big)}{\big((\omega - E_{\vec{k}})^2 - c^2(\vec{q} - \vec{k})^2)^{1-\eta/2}} \label{eq: integral}
\end{equation}
where $E_{\vec{k}}$ is the energy of the $f$-fermions relative to the Fermi energy. The only contribution comes from a region in k space with $E_{\vec{k}} < \omega$ and $c^2 (\vec{q} - \vec{k})^2 < (\omega - E_{\vec{k}})^2$. These conditions must be simultaneously satisfied to obtain a non-zero spectral function. In the limit $\omega \rightarrow 0$, this means that $A(\vec{q}, \omega)$ is only non-zero if $E_{\vec{q}} \approx 0$ i.e. if $\vec{q}$ lies near the critical Fermi surface. In this limit, the only contribution comes from a window with $\vec{k} \approx \vec{q}$, of width $\omega/v_f$ perpendicular to the Fermi surface, and breadth $(\omega/c)^{d-1}$ along the Fermi surface. The spectral function for $\vec{q}$ precisely on the critical Fermi surface then behaves as
\begin{equation}
A(\vec{q} \mathrm{\,\,on \,\, FS}, \omega \to 0) \sim \omega^{d - 2+ \eta} 
\label{eq:Aon}\end{equation}
More generally, for $\vec{q}$ slightly off the Fermi surface,
\begin{equation}
A(\vec{q}, \omega) \approx \omega^{d - 2+ \eta} g\left(\frac{\omega}{v_f q_\perp}\right) \label{eq:Ascal}
\end{equation}
where $q_{\perp}$ is the momentum deviation from the Fermi surface. The function $g(x) \to const$ for $x \to \infty$ and $g(x) = 0$ for $-c/v_f < x < \mathrm{min}(c/v_f, 1)$. Thus, in the present approximation,
\be A(\vec{q} \mathrm{\,\,off \,\, FS}, \omega \to 0) = 0 \label{eq:Aoff}\ee
Contrasting Eqs.~\ref{eq:Aon}, \ref{eq:Aoff}, we see that the Fermi surface remains sharply defined at the quantum critical point. Integrating the spectral function over momenta, we obtain the local tunneling density of states at the transition,
\begin{equation}
N(\omega) \sim \omega^{d - 1 + \eta}
\end{equation}
We must note that our calculation has treated the $f$ fermions as non-interacting. Once interactions between the $f$'s are included, the electron spectral weight for momenta off the Fermi surface will no longer have a hard gap as in Eq.~\ref{eq:Aoff}. However, as we now show, the low frequency spectral weight off the Fermi surface is suppressed compared to the spectral weight on the Fermi surface by powers of $\omega$, so that a sharp notion of the Fermi surface persists.


As noted in section \ref{sec:inter} sufficiently weak interactions between the $f$ fermions drive them into a Fermi-liquid phase, where their spectal function takes the form
\begin{equation}
A_f(\omega, \vec{k}) = \frac{K \omega^2}{(\omega - \epsilon_{\vec{k}})^2 + K^2 \omega^4}
\end{equation}
 where $\epsilon_{k}$ is the renormalized dispersion, $K$ is a constant, and we assume that the imaginary part of the self energy $\Sigma$ scales as $\Im\big(\Sigma(\omega\rightarrow 0)\big) = K\omega^2$. (Stricly speaking, $K$ has a logarithmic dependence on $k_\perp$ in $d=2$, this, however, plays no role in our discussion below.) $A_f$ must be convolved with the Ising spectral function according to (\ref{eq: spectral function2}) to obtain the spectral function for electrons. At the critical point, the electron spectral function takes the form
\begin{widetext}
\begin{eqnarray}
A(\vec{q},\omega) &=& \int \frac{d^dk}{(2\pi)^d} \int_0^{\omega} d\Omega \frac{\Theta\big((\omega-\Omega)^2 - c^2 |\vec{k}-\vec{q}|^2\big)}{\big((\omega-\Omega)^2 - c^2 |\vec{k}-\vec{q}|^2\big)^{1-\eta/2}}\frac{K \Omega^2}{(\Omega - \epsilon_{\vec{k}})^2 + K^2 \Omega^4}\nonumber\\
&=& \omega^{-1+\eta}  \int_0^{1} dx \int \frac{d^dk}{(2\pi)^d} \frac{\Theta\big((1-x)^2 - (c/\omega)^2 |\vec{k}-\vec{q}|^2\big)}{\big((1-x)^2 - (c/\omega)^2 |\vec{k}-\vec{q}|^2\big)^{1-\eta/2}}\frac{K x^2}{(x - \epsilon_{\vec{k}}/\omega)^2 + K^2 \omega^2 x^2}
\end{eqnarray}
\end{widetext}
where we define $\Omega = \omega x$. We now consider the scaling limit $\omega \rightarrow 0$. In this limit, the theta function restricts the momentum integral to a hypersphere of volume $(\omega/c)^d$ centered at $\vec{q}$. Thus, we obtain
\begin{equation}
A(\vec{q},\omega\rightarrow0) \sim \omega^{-1+d + \eta}  \int_0^{1} dx \frac{K x^2}{(x - \epsilon_{\vec{q}}/\omega)^2 + K^2 \omega^2 x^4}
\end{equation}
This has very different behavior for $\epsilon_{\vec{q}} = 0$ (i.e. $\vec{q}$ on the Fermi surface) and for $\epsilon_{\vec{q}} \neq 0$. For $\epsilon_{\vec{q}} \neq 0$, the second term in the denominator can be neglected in the scaling limit $\omega \ll \epsilon_{\vec{q}}$. As a result, the spectral function takes the form
\begin{eqnarray}
A(\vec{q}\text{ off FS},\omega\rightarrow0) \sim \omega^{d +1 + \eta}
\label{eq:Aoff2}
\end{eqnarray}
Meanwhile, for the special case when $\vec{q}$ is on the Fermi surface and $\epsilon_{\vec{q}}=0$, we obtain as in Eq.~\ref{eq:Aon},
\begin{eqnarray}
A(\vec{q}\text{ on FS},\omega\rightarrow0) \sim \omega^{d -2 + \eta}
\end{eqnarray}
The sharp difference in the scaling limit of the spectral function indicates that the critical Fermi surface remains sharply defined, even when the $f$- fermions are in a Fermi liquid state. 

\subsection{Effects beyond mean field}
\label{ebmf}
In the previous section we discussed the evolution of the spectral function in a `mean-field' approximation wherein the slave boson and Ising fermion sectors were assumed to be decoupled. We now discuss the robustness of this mean field approximation. The Ising transition in the slave spin sector is described by a continuum relativistic massless $\phi^4$ theory \cite{Sachdev}. However, this $\phi^4$ theory can couple to the fermion sector, as discussed in Ref. \onlinecite{Morinari}. In particular, the `energy' operator $O = \phi^2$ can couple to particle-hole excitations of the $f$ fermions. Such a coupling is potentially dangerous, due to the existence of gapless particle-hole excitations close to the Fermi surface. It is convenient to first discuss the situation for continuum realizations of the Orthogonal Metal. In the continuum the most dangerous coupling is between the Ising energy operator and the fermion density. Integrating out the $f$ fermions introduces `Landau damping' to the slave spin sector, adding a term to the Hamiltonian which takes the form \cite{Morinari}
\begin{equation}
\delta S = v \int_{\omega, \vec{q}} \Pi(\omega, q) |O(\vec{q},\omega)|^2. \label{eq: Landau damping'}
\end{equation}
where $\Pi(\omega, q)$ is the density-density correlator of the fermions.  If this `new' term is irrelevant (in the renormalisation group sense), then the mean field treatment that ignores the coupling of the slave spins to the $f$ fermions is robust, whereas if it is relevant, then it may change the universality class of the transition. Note that higher composites of $O$ induced by integrating out the $f$ fermions are less relevant in the RG sense than the operator (\ref{eq: Landau damping'}) and so will not be discussed further.
 
If the electron-electron interaction is short ranged, $\Pi(\omega, \vec{q}) \sim C(\omega/q)$ for $(\omega, q) \to 0$. For future reference, we note that the function $C$ behaves as $C(\omega/q) \approx \kappa_0 + c \frac{|\omega|}{q} + \ldots$ for $\omega/q \to 0$, although at present we are interested in the limit $\omega/q \sim O(1)$ dictated by the $z=1$ dynamics of the Ising transition, where $C \sim O(1)$. In $D$ space-time dimensions, the operator $O$ has scaling dimension $D-1/\nu$ at the critical point, where $\nu$ is the correlation length exponent. The frequency dependent  Landau damping term is then relevant if $\nu < \frac{2}{D}$. For the $2+1$ dimensional Ising transition $D=3$ and $\nu = 0.63$ so that the coupling of the critical Ising sector to the Fermi surface is weakly relevant. For the $3+1$ dimensional Ising model, $D = 4$, $\nu = 1/2$ and this same coupling is marginal by power counting. Analysis of renormalization group flows at one loop level leads to flows to strong coupling both in $d = 3$ and just below $3$ dimensions so that no confident conclusions can be drawn. Thus, both in two and three spatial dimensions with short ranged interactions the Ising transition is destabilized but we do not know whether it is replaced by a second order transition in a new universality class or a first order transition. In either case, the near marginality  of  the Fermi surface coupling means that there may be a long crossover critical regime where the physics is controlled by the decoupled fixed point described above. 

The presence of long ranged Coulomb interactions dramatically simplifies the physics. 
The key point is that the $f$ fermions are electrically charged, and the Coulomb interaction gaps out the density fluctuations (case $A$ of Ref. \onlinecite{Morinari}). Formally, the density-density correlator is given  by 
\begin{equation}
\Pi(\omega, q) = \frac{\Pi_0(\omega, q)}{1 + V(q) \Pi_0(\omega, q)} \label{eq: RPA}
\end{equation}
where $\Pi_0(\omega, q)$ is the irreducible density correlator and $V(q)$ is the Coulomb interaction, taking the form $V(q) = e^2/q$ in two spatial dimensions. In the scaling limit $\omega \rightarrow 0$, $q \rightarrow 0$, $\omega/q$ finite, we have $\Pi_0$ finite and $V(q) \rightarrow \infty$. It is now convenient to rewrite (\ref{eq: RPA}) as 
\begin{equation}
\Pi(\omega, q) = \frac{V^{-1}(q)}{1 + V^{-1}(q) \Pi_0^{-1}(\omega, q)} \label{eq: RPA'}.
\end{equation}
Taylor expanding (\ref{eq: RPA'}) in small $V^{-1}(q)$, we  find that the `Landau damping' arising from coupling to a Fermi surface of charged particles, evaluated in the scaling limit, takes the form
\begin{equation}
\delta S = v \int_{\omega, \vec{q}} q \left(1 - \frac{q}{e^2} \Pi^{-1}_0(\omega/q)\right) |O(\vec{q},\omega)|^2. \label{eq: Landau damping}
\end{equation}
We now find that $v$ is irrelevant if $\nu > 2/(D+1)$, but is relevant if $\nu < 2/(D+1)$. For the Ising model in $2+1$ dimensions, $\nu = 0.63 > 1/2$, indicating that the modified Landau damping term (\ref{eq: Landau damping}) is irrelevant at tree level. As a result, the slave spin and orthogonal fermion sectors decouple at low energies, and the mean field treatment is robust. Likewise, in $3+1$ dimensions, $v(q) = e^2/q^2$ so the leading fermion induced term in the Lagrangian for $\phi$ becomes $\phi^2 \nabla^2 \phi^2$, which is strongly irrelevant.

Very different results are obtained if we consider a system without continuous rotation symmetry (e.g. a generic lattice model). In this case, the Fermi surface can undergo long wavelength deformations which preserve the full lattice symmetry, and also do not change the volume enclosed by the Fermi surface. Such higher moment deformations of the Fermi surface can also couple to the slave spin sector, and crucially are not gapped out by the Coulomb interaction. They introduce a `conventional' Landau damping term into the action
\begin{equation}
\delta S = v \int_{\omega, \vec{q}} C(\omega/q) |O(\vec{q},\omega)|^2. \label{eq: Landau damping2}
\end{equation}
We are then back in the same situation as with the short ranged case in the continuum considered above, and thus the decoupled fixed point is destabilized for $d =2$, and may be marginally destabilized for $d = 3$.

We have thus reached the conclusion that in generic lattice models the Ising transition from the Fermi liquid to the OM is destabilized by coupling between the slave spin and $f$ fermion sectors. The solvable models described in the next section are fine tuned in that there is no coupling at all between the Fermi surface and the slave spin sector. Perturbing about the solvable limit will then lead to a weakly relevant flow to a different universality class (or possibly a first order transition) but with a wide intermediate region controlled by the transition in the solvable limit. 

\subsection{Phase transition in the $Z_4$ slave-spin model}
Is it possible to have a controlled theory of a sharply defined critical Fermi surface in a lattice model without having to fine tune the coupling to the Fermi surface to zero? The answer is yes. To see this consider a generalized OM where the slave spin is a $Z_4$ rather than $Z_2$ variable. The slave-spin ordering transition is described by a theory,
\be L = \Phi^* \der_\tau \Phi + \frac{1}{2m_\Phi} |\nabla \Phi|^2 + t |\Phi|^2 + \frac{u}{4} |\Phi|^4 + \frac{v}{4!} (\Phi^4 + (\Phi^*)^4) \label{LZ4}\ee

Here $\Phi$ is a complex scalar field corresponding to a coarse-grained $Z_4$ spin and transforming as $\Phi \to i \Phi$ under the ${Z}_4$ symmetry. Note that a first order time derivative term $\Phi^* \der_\tau \Phi$ is allowed by symmetry and will be generically present (see appendix \ref{app:PH} for further discussion). Thus, at mean-field level, the theory (\ref{LZ4}) has a dynamical exponent $z = 2$. This should be contrasted with the case of a ${Z}_2$ slave-spin discussed in the rest of the paper, where the slave-spin sector has an emergent Lorentz symmetry and a dynamical critical exponent $z = 1$.

We can repeat the calculations of section \ref{sec:AMF} to find how the electron spectral function evolves across the slave-spin ordering transition. At mean-field level we find that the quasiparticle residue $Z \sim \langle \Phi \rangle^2 \sim (t_c - t)$ in the Fermi-liquid phase and the electron gap $\Delta \sim (t-t_c)$ in the orthogonal metal phase. At the transition, using $A_{spin}(\omega, \vec{q}) \sim \delta(\Omega - \vec{q}^2/(2 m_\Phi))$ we find,
\be A(q_\perp, \Omega) \sim \left(\Omega - \frac{q^2_\perp}{2 m_\Phi}\right)^{(d-1)/2} \theta\left(\Omega - \frac{q^2_\perp}{2 m_\Phi}\right)\label{AZ4}\ee
Thus, the Fermi-surface is critically destroyed at the slave-spin ordering transition. It is amusing to note that the critical spectral function is not invariant under the dynamical particle hole transformation $\Omega -> -\Omega, q_\perp -> -q_\perp$ unlike what happens in a Landau fermi liquid. The possibility of such a  dynamical particle-hole asymmetry in correlated non-fermi liquid metals has recently been emphasized in \onlinecite{Casey} and \onlinecite{Shastry}.

Proceeding beyond mean-field in the analysis of the theory (\ref{LZ4}), the quartic couplings $u$ and $v$ are irrelevant in $d = 3$ and marginal at tree-level in $d = 2$. In the latter case, one loop calculations performed in appendix \ref{app:Z4RG} demonstrate that there exists a finite region in the $(u,v)$ plane where the RG trajectrories flow to the non-interacting fixed point $(u,v) = (0,0)$. Thus, the mean-field treatment of the theory (\ref{LZ4}) is also valid in $d = 2$ up to logarithmically suppressed corrections.

Next, we consider the coupling of the slave-spin sector (\ref{LZ4}) to the $f$-Fermi surface. Again the most relevant coupling is between the operator $O = |\Phi|^2$ and $f$-bilinears describing deformations of the Fermi surface. Integrating the $f$'s out, we again generate a ``Landau-damping" term (\ref{eq: Landau damping'}). However, since the slave-spin sector is now characterized by dynamical scaling $\omega \sim q^2 \ll q$, it is permissible to expand the function $\Pi(\omega, q) = C(\omega/q)$ in $\omega/q$,
{\begin{equation}
\delta S = v \int_{\omega, \vec{q}} \left(\kappa_0 + c \frac{|\omega|}{q} + \ldots\right) |O(\vec{q},\omega)|^2. \label{eq: Landau damping3}
\end{equation}}
{The $\kappa_0$ term in Eq.~\ref{eq: Landau damping3} simply renormalizes the coupling constant $u$ in Eq.~\ref{LZ4}, while the first subleading term $|\omega|/q \sim O(q)$ is irrelevant under RG. Therefore, the coupling of the slave-spin theory (\ref{LZ4}) to the $f$ Fermi surface does not change the universality class of the transition. This transition then provides a simple example of a continuous destruction of the electron Fermi surface that is accompanied by a critical Fermi surface.} {In the appendix \ref{app:PH} we discuss certain distinctions between the present transition and transitions out of Mott-Insulating states.} 

\section{Exactly soluble models}
{We now present two exactly soluble models that realize the Orthogonal Metal phase and its phase transition to the conventional Fermi liquid. The second model will have an extra feature, namely, a global Ising symmetry will be spontaneously broken upon entering the Fermi liquid phase; this will have a non-trivial effect on the critical Fermi surface at the phase transition. As a stepping stone for our construction, we will use a model of quantum Ising spins that possesses a confinement-deconfinement transition between a topologically ordered ${Z}_2$ spin-liquid phase and a conventional paramagnetic phase.}

\subsection{A model for the confinement-deconfinement transition}
\label{sec:confdeconf}
{Consider a model of quantum Ising spins placed on the links $\langle rr'\rangle$ of a square lattice, and coupled according to}

\begin{equation}
{H = -J \sum_{\langle rr'\rangle} \sigma^z_{rr'}  - h \sum_r \prod_{r'_r} \sigma^x_{rr'} \label{eq: sqexact}}
\end{equation}
with $J>0$ and $h>0$, where the product over $r'_r$ denotes a product over all four links connected to the site $r$. One can introduce an infinite number of operators $\Phi^p_r$ that commute with the Hamiltonian, which are defined by 
\begin{equation}
\Phi^p_{r_1r_2r_3r_4} = \sigma^z_{r_1 r_2} \sigma^z_{r_2 r_3}\sigma^z_{r_3 r_4} \sigma^z_{r_4 r_1} \label{eq:Phip}
\end{equation}
where $\{r_1 r_2 r_3 r_4\}$ are the four corners of a square plaquet. Since the symmetry operators $\Phi^p$ commute with each other as well as with the Hamiltonian, we can label the eigenstates of the Hamiltonian by their $\Phi^p$ eigenvalues. The ground state always has $\Phi^p_{r_1 r_2 r_3 r_4} = 1$ for all plaquets, so we introduce new `fractionalized' Ising spins $\tau$ that live on the sites of the square lattice, which obey
\begin{equation}
 \tau^x_r \tau^x_{r'} = \sigma^z_{rr'} \qquad \prod_{r'_r} \sigma^x_{rr'} = \tau^z_r \label{eq:tautransf}
\end{equation}
The Hamiltonian, rewritten in terms of the spins $\tau$, takes the form
\begin{equation}
{H = -J \sum_{\langle rr'\rangle} \tau^x_{r} \tau^x_{r'} - h \sum_r  \tau^z_{r}}
\label{HTFI}\end{equation}
{We recongnize Eq.~\ref{HTFI} as the Hamiltonian of the transverse field Ising model. As we tune $h/J$, the Ising model undergoes an Ising phase transition. The ordered phase of the $\tau^x$ spins corresponds to the conventional ``confined" paramagnetic phase, while the disordered phase corresponds to the ``topologically" ordered ${Z}_2$ spin-liquid.}


\subsection{Soluble model for the Orthogonal Metal}
\label{sec:OMmodel1}
We now introduce electrons which hop on the square lattice according to 
\begin{equation}
H_t = - t \sum_{\langle r r' \rangle} \left(c^{\dag}_r \sigma^z_{rr'} c_{r'} + h.c.\right) - \mu \sum_r c^{\dag}_rc_r \label{eq:Ht}
\end{equation}
{We also modify the Hamiltonian of the quantum Ising spins to be,}
\begin{equation}
{H_\sigma = -J \sum_{\langle rr'\rangle} \sigma^z_{rr'}  - h \sum_r (-1)^{c^{\dagger}_r c_r} \prod_{r'_r} \sigma^x_{rr'}  \label{eq: sqexact2}}
\end{equation}
{We will study the coupled Hamiltonian $H = H_t + H_\sigma$.}

Changing to the $\tau$ representation, and defining 
{\be f_r = \tau^x_r c_r\label{eq:fdef}\ee}
we obtain 
{
\bea
H_t &=&  -t \sum_{\langle rr' \rangle} \left(f^{\dag}_r f_{r'} + h.c.\right) - \mu \sum_r f^{\dag}_r f_r\label{eq:Htf}\\
H_\sigma &=& -J \sum_{\langle rr'\rangle} \tau^x_{r} \tau^x_{r'} - h \sum_r  \tau^z_{r} (-1)^{f^\dagger_r f_r}
\eea
}
{A further transformation},
\be {\tilde{\tau}^z_r = \tau^z_{r} (-1)^{f^\dagger_r f_r}, \quad \tilde{\tau}^x_r = \tau^x_r}\label{eq:tautilde}\ee
{gives}
\be {H_\sigma = -J \sum_{\langle rr'\rangle} \tilde{\tau}^x_{r} \tilde{\tau}^x_{r'} - h \sum_r  \tilde{\tau}^z_{r}} \label{eq:TFIM2}\ee
{and ensures that the $f$-operators commute with the $\tilde{\tau}$ operators, which satisfy the conventional Pauli algebra. Thus, $H_t$ describes free fermions $f$, which are completely decoupled from the $\tilde{\tau}$ spins. The latter are still governed by the transverse field Ising model (\ref{eq:TFIM2}), which undergoes a phase transition as a function of $h/J$.}

The fermions $f$ carry all the quantum numbers of the electron (in particular the electric charge), and the $f$ fermions are manifestly in a metallic phase. Thus, the system is conducting. However, it is not necessarily a Fermi liquid. In the ordered phase of the $\tilde{\tau}$ spins, $Z = \langle \tau^x_r\rangle^2 \neq 0$, and the $f$ fermions have non-vanishing overlap $Z$ with the free electrons. This phase corresponds to a Fermi liquid. Meanwhile, in the disordered phase of the spins, $Z =\langle \tau^x_r \rangle^2 = 0$, and the $f$ fermions are orthogonal to free electrons (Orthogonal Metal). 

{The exactly soluble model presented above thus provides an explicit realization of the `Orthogonal Metal' phase and its transition into the Fermi-liquid phase. Since the $f$ and the $\tilde{\tau}$ sectors are decoupled in the present model, the mean-field results of section \ref{sec:AMF} hold exactly. In particular, at the transition a sharp critical Fermi surface is realized.} 

\subsection{Soluble model for an Ising$^*$ non-Landau quantum critical point and its metallic generalization }
{In this section we modify the model considered above so that it has a global Ising symmetry. In the absence of fermions, the new model will describe a topologically ordered paramagnetic phase, a conventional Ising ferromagnetic phase where the global symmetry is spontaneously broken and a non-Landau quantum critical point separating them. Following the usual terminology\cite{LGWstar} we will refer to this as an Ising$^*$ transition. Once the fermions are introduced, the topologically ordered paramagnetic phase will turn into an Orthogonal Metal and the conventional Ising ferromagnetic phase into a ferromagnetically ordered Fermi-liquid. At the transition, a critical Fermi-surface will be realized with a volume that violates the conventional Luttinger count.}
%

{We consider the Hamiltonian, $H = H_t + H_\sigma$ where $H_t$ is still given by Eq.~\ref{eq:Ht} and $H_\sigma$ is},
\begin{equation}
{H_\sigma = -J \sum_{r} \sum_{\mu = \pm \hat{x},\nu = \pm \hat{y}} \sigma^z_{r,r+\mu} \sigma^z_{r, r+\nu} -h \sum_r (-1)^{c^{\dagger}_r c_r} \prod_{r'_r} \sigma^x_{rr'}} \label{eq: sqexact3}
\end{equation}
{Here we have defined the model with fermionic degrees of freedom from the outset, however, if desired, one can consider the sector with zero fermion number and obtain a model of quantum Ising spins alone. Note that we have modified the $J$ term in Eq.~\ref{eq: sqexact2} to couple nearest neighbour spins. As a result, the new model has a global Ising symmetry,
\be \sigma^z_{rr'} \to -\sigma^z_{rr'},\quad c_r \to \epsilon_r c_r \label{eq:IsingSymm}\ee
Here $\epsilon_r = 1$ for $r$ on one sublattice of the square lattice, and $\epsilon_r = -1$ for $r$ on the other sublattice. 
As before, the operators $\Phi^p$ (\ref{eq:Phip}) commute with the Hamiltonian. We again work in the ground state sector $\Phi^p = 1$ and use the transformation to the $\tilde{\tau}$ and $f$ variables in Eqs.~\ref{eq:tautransf},\ref{eq:fdef},\ref{eq:tautilde} to obtain},
\be 
H_\sigma = -2 J \sum_{\langle\langle rr'\rangle\rangle} \tilde{\tau}^x_{r} \tilde{\tau}^x_{r'} - h \sum_r  \tilde{\tau}^z_{r} \label{eq:Hsigma2}\ee
where $\langle \langle \rangle \rangle$ denotes second neighbors. The Hamiltonian $H_t$ in the transformed variables is still given by Eq.~\ref{eq:Htf}.

The Hamiltonian $H_\sigma$ (\ref{eq:Hsigma2}) now takes the form of two decoupled transverse field Ising models on the two sublattices of the square lattice. As we tune $h/J$, the Ising models undergo an Ising phase transition. The `physical' order parameter that transforms under the global Ising symmetry (\ref{eq:IsingSymm}) is  $\langle \sigma^z_{rr'} \rangle = \langle \tilde{\tau}^x_r \rangle \langle \tilde{\tau}^x_{r'} \rangle $. At the critical point the correlations of the physical Ising spin satisfy 
\begin{equation}
\langle \sigma^z_{r_1r_1'} \sigma^z_{r_2r_2'} \rangle = \langle \tilde{\tau}^x_{r_1} \tilde{\tau}^x_{r_2} \rangle \langle \tilde{\tau}^x_{r_1'} \tilde{\tau}^x_{r_2'} \rangle
\end{equation}
As each fractionalized $\tilde{\tau}^x$ undergoes the usual Ising transition it follows that the physical Ising order parameter has anomalous dimension $\eta_{\sigma} = 1+ 2\eta$ where 
$\eta$ is the anomalous dimension of the Ising order parameter at the usual 3D Ising transition. The phase transition in this solvable model is thus a very simple example of a non-Landau quantum critical point\cite{LGWstar,maissam} between a conventional phase with Landau order and a topologically ordered phase. 

{Note that the ferromagnetic transition takes place both at zero and finite fermion density. Let us discuss the nature of the phases at finite density by examining the electron Green's function. Since the $f$ and $\tilde{\tau}$ sectors are decoupled, we have,
\bea
G(\vec{r}-\vec{r'}, \tau -\tau') &=& \langle c_{\vec{r}}(\tau) c^{\dag}_{\vec{r'}}(\tau')\rangle \nn
&=& \langle \tilde{\tau}^x_{\vec{r}} (\tau) \tilde{\tau}^x_{\vec{r'}}(\tau')\rangle G_f(\vec{r}-\vec{r'}, \tau - \tau')\nonumber\\
\label{eq:Gprod}\eea
The $f$-fermions are governed by a free fermion hopping Hamiltonian, and thus have some well defined Fermi surface in momentum space obeying the Luttinger count. 
In the paramagnetic phase of the slave spins, the slave spin correlation function is gapped, and so the electrons do not have a Fermi surface. We, thus, identify this phase as an Orthogonal Metal. 
On the other hand, in the ferromagnetic phase of the slave spins, the slave spin correlator is non-vanishing at long distances/times, and so this phase is a Fermi-liquid. What is the shape of the Fermi-surface in this phase? To answer this question note that the ground state in the ferromagnetic phase is two-fold degenerate, spontaneously breaking the global symmetry (\ref{eq:IsingSymm}). The two ground states with opposite expectation values of the order parameter $\langle \sigma^z_{rr'} \rangle$ correspond to the $\tilde{\tau}^x$ spins on the two decoupled sublattices oriented parallel or antiparallel to each other. The $\tilde{\tau}^x$ correlation function in the two ground states, therefore, in the long distance/time limit takes the form,
\bea \langle \tilde{\tau}^x_\vec{r}(\tau) \tilde{\tau}^x_{\vec{r}'}(\tau')\rangle &\sim& \,\,\,\langle \tilde{\tau}^x\rangle^2, \quad \vec{r}, \vec{r}' \,\,\mathrm{on\,\,same\,\,sublattice} \nn &=& \pm \langle \tilde{\tau}^x\rangle^2, \quad \vec{r}, \vec{r}' \,\,\mathrm{on\,\,different\,\,sublattices}\nonumber\eea
Hence, $\langle \tilde{\tau}^x_\vec{r}(\tau) \tilde{\tau}^x_{\vec{r}'}(\tau')\rangle \sim \langle \tilde{\tau}^x\rangle^2$ in one ground state and $\langle \tilde{\tau}^x_\vec{r}(\tau) \tilde{\tau}^x_{\vec{r}'}(\tau')\rangle \sim \langle \tilde{\tau}^x\rangle^2 e^{i \vec{Q} \cdot (\vec{r}-\vec{r}')}$ in the other ground state. Here $\vec{Q} = (\pi, \pi)$. Therefore, the electron Green's function, $G(\vec{k}, \omega) \sim \langle \tilde{\tau}^x\rangle^2 G_f(\vec{k}, \omega)$ in one ground state and $G(\vec{k}, \omega) \sim \langle \tilde{\tau}^x\rangle^2 G_f(\vec{k} + \vec{Q}, \omega)$ in the other. So the Fermi-surface in one ground state is the same as the $f$ Fermi-surface, and the Fermi-surface in the other ground state is given by the $f$ Fermi-surface shifted by $\vec{Q}$. This is a direct consequence of the global Ising symmetry (\ref{eq:IsingSymm}), which maps the two ground states into each other and transforms $G(\vec{k},\omega) \to G(\vec{k} + \vec{Q},\omega)$. Note that the Fermi-surfaces of both ground states satisfy the Luttinger theorem.}

{What happens to the Fermi-surface at the critical point? As we approach the transition from the ferromagnetically ordered side, the quasiparticle residue  $Z = \langle \tilde{\tau}^x\rangle^2$ continuously goes to zero on Fermi-surfaces of both ground states. At the transition, the Ising symmetry (\ref{eq:IsingSymm}) is restored, so the ground state is unique and the electron Green's function satisfies $G(\vec{k}) = G(\vec{k} + \vec{Q})$. Hence, we expect the electron spectral function at the transition to inherit Fermi-surfaces of both ground states of the ferromagnetic phase, albeit as critical Fermi-surfaces. To see that this is, indeed, the case, note that at the transition,
\bea \langle \tilde{\tau}^x_\vec{r}(\tau) \tilde{\tau}^x_{\vec{r}'}(\tau')\rangle &=& \,\,\, G_{spin}(\vec{r}-\vec{r}',\tau - \tau'),\nn 
&& \quad \vec{r}, \vec{r}' \,\,\mathrm{on\,\,same\,\,sublattice}; \nn \langle \tilde{\tau}^x_\vec{r}(\tau) \tilde{\tau}^x_{\vec{r}'}(\tau')\rangle &=& 0, \quad \vec{r}, \vec{r}' \,\,\mathrm{on\,\,different\,\,sublattices}\nonumber\eea
Here, $G_{spin}(\vec{r}, \tau)$ is the correlation function of the transverse field Ising model on just one sublattice of the square lattice with the assymptotic behavior $G_{spin}(\vec{r}, \tau) \sim 1/(r^2 + c^2 \tau^2)^{(d-2 + \eta)/2}$ at the critical point. Written more succintly, we have 
\be \langle \tilde{\tau}^x_\vec{r}(\tau) \tilde{\tau}^x_{\vec{r}'}(\tau')\rangle = \frac{1}{2}(1+e^{i \vec{Q} \cdot (\vec{r}-\vec{r}')}) G_{spin}(\vec{r}-\vec{r'}, \tau-\tau')\nonumber\ee
 Therefore, from Eq.~\ref{eq:Gprod}, $G(\vec{k}, \omega) = G(\vec{k}+\vec{Q}, \omega)$ as required by the global symmetry. Moreover, using the analysis of section \ref{sec:AMF}, the electron spectral function at the phase transition displays two sharply defined critical Fermi-surfaces: one at the $f$ Fermi-surface, and one obtained by translating the $f$ Fermi-surface by $\vec{Q}$. We call these `mirror' critical Fermi-surfaces. Note that the total volume enclosed by the two mirror Fermi-surfaces violates the Luttinger count by a factor of two. However, the state realized at the critical point is not a Fermi-liquid, so there is no reason for Luttinger's theorem to be satisfied.}

{Before we conclude this section, we stress that due to the additional global symmetry (\ref{eq:IsingSymm}), the critical point of the present model is in a different universality class compared to the simplest Orthogonal Metal to Fermi-liquid transition studied in the rest of this paper and examplified by the exactly solvable model of section \ref{sec:OMmodel1}. In particular, the appearance of the mirror critical Fermi-surfaces at the critical point is a direct consequence of the symmetry (\ref{eq:IsingSymm}). Once this symmetry is explicitly broken, only one critical Fermi-surface with a volume obeying Luttinger theorem will be realized at the transition, as is directly seen in the model of section \ref{sec:OMmodel1}.}

\subsection{Orthogonal metal in one spatial dimension}
A one dimensional model with a $Z_2$ fractionalized phase and an Ising transition is easy to write down: a suitable Hamiltonian is 
{\bea
H &=& - \sum_{i}  \left(J \sigma^x_{i-1, i} \sigma^x_{i, i+1} (-1)^{c^{\dagger}_i c_i}  + h \sigma^z_{i, i+1}\right) \nn
 &-& t \sum_i \left(c^{\dag}_i \sigma^z_{i, i+1} c_{i+1} + h.c.\right)
\eea
}where the $c^{\dag}$ operators create electrons, and the $\sigma$ operators act on Ising spins that live on the links of a $1D$ chain. 
We now make use of the well known self-duality of the one dimensional Ising model, defining $\sigma^z_{i, i+1} = \tau^x_i \tau^x_{i+1}$ and $\sigma^x_{i-1, i} = \prod_{j<i} \tau^z_i$. The Hamiltonian then becomes

\bea
H &=& - \sum_{i} \left(h \tau^x_{i} \tau^x_{i+1} + J \tau^z_{i} (-1)^{c^{\dagger}_i c_i}\right)\nn
&-& t \sum_i \left(c^{\dag}_i \tau^x_{i} \tau^x_{i+1} c_{i+1} + h.c.\right)
\eea
Defining new fermions $f_i = \tau^x_i c_i$ (equivalently, $c_i = \tau^x_i f_i$), and making an additional change of variables,
\be \tilde{\tau}^z_i = \tau^z_i (-1)^{f^{\dagger}_i f_i}, \quad \tilde{\tau}^x_i = \tau^x_i\ee
we see that the fermion and spin sectors decouple, and the fermion sector becomes simply a free fermion theory, while the spin sector becomes a transverse field Ising model. Since the $f$ fermions carry the electric charge, and are described by a free fermion Hamiltonian, it follows that this Hamiltonian always exhibits a non-vanishing electrical conductivity. Thus, the system is always in a metallic phase. However, the charge carriers are not electrons - they are $f$ fermions, which are non-local combinations of electrons and $\sigma$ spins. 

To understand the nature of the charge carriers, note that $f_i = \tau^x_i c_i$, where $\tau^x_i = \prod_{j<i} \sigma^z_{j j+1}$ creates a domain wall in the $\sigma$ spin ferromagnet. In the paramagnetic phase of the original Ising model, $h \gg J$, the domain walls are condensed, and $f_i  \propto c_i$. This phase corresponds to the `Fermi liquid' (short range four fermion interactions would drive it into a Luttinger liquid state \cite{Giamarchi}). Meanwhile, in the ferromagnetic phase of the original Ising model, $h \ll J$, the domain wall is a (gapped) topological defect. In this phase, the $f$ fermion charge carriers are orthogonal to the electrons, and correspond to a bound state of a domain wall and an electron. The model is constructed so that this bound state can hop freely, producing an Orthogonal Metal. 

\section{Realizing an Orthogonal Metal} 
We have thus far demonstrated the existence of the novel Orthogonal Metal phase and described its properties in some detail.   We have presented several solvable models that realize such a phase, however, none of these models is particularly realistic as a description of experimental systems. In what more realistic setting could an Orthogonal Metal phase be realized? 

The structure of the Orthogonal Metal wavefunction provides some clues as to when this phase might occur. Briefly, the Orthogonal Metal is similar to the Fermi liquid, except that there are strong correlations between electron pairs. At the same time, there is no phase coherence between electron pairs (no superconductivity). This suggests that a promising Hamiltonian for potentially realizing an Orthogonal Metal should contain strong electron-electron repulsion (to prevent superconductivity), but should also contain strong pair hopping, to favor pair correlations. For example, the half filled triangular lattice with a single orbital species of spinful electrons governed by the Hamiltonian 
\bea
H &=&- \sum_{\langle ij \rangle} \left(\big(\sum_{\sigma} t_1 c^{\dag}_{i \sigma} c_{j \sigma}\big) + t_2 c^{\dag}_{i \uparrow} c^{\dag}_{i \downarrow} c_{j \downarrow}c_{j \uparrow} + h.c.\right) \nn
&+& U \sum_i c^{\dag}_{i \uparrow} c^{\dag}_{i \downarrow} c_{i \downarrow}c_{i \uparrow} \label{eq: triangular hamiltonian},
\eea
may exhibit an Orthogonal Metal phase. Applying the slave boson representation \ref{eq: slave bosons} and the constraint \ref{eq: constraint2} and decoupling the slave bosons from the fermions in a mean field approximation, we obtain in the boson sector the Hamiltonian
\begin{equation}
H_b = -\sum_{\langle i j \rangle} \left(t'_1 b^{\dag}_i b_j + t'_2 b^{\dag}_i b^{\dag}_i b_j b_j + h.c.\right) + U \sum_i b^{\dag}_i b_i (b^{\dag}_i b_i -1)
\end{equation}
where ${t_1' =  2 t_1\langle d^{\dagger}_{i\sigma} d_{j\sigma} \rangle}$ and ${t_2' = t_2 \langle d^{\dagger}_{i\uparrow} d^{\dagger}_{i\downarrow} d_{j\downarrow} d_{j\uparrow} \rangle}$. In the limit $t_2 \gg U$, we expect superfluidity of boson pairs. Meanwhile, in the limit $U \gg t_1$ we expect no superfluidity of individual bosons. This is precisely the condition for realization of an Orthogonal Metal. Thus, we expect that the Hamiltonian \ref{eq: triangular hamiltonian}, defined on the triangular lattice at half filling for a single species of (spinful) fermions, may possess an Orthogonal Metal phase at $t_2 \gg U \gg t_1$. Such a Hamiltonian could conceivably be produced by strong electron-phonon coupling \cite{Assad}. 

We also note that since the original discussion of multiband Hubbard models of Ref. \onlinecite{Georges}, many other studies have emerged where the slave spin mean field has been employed and found regions of parameter space where the slave spins are disordered. We have emphasized that these should be interpreted as Orthogonal Metals or Orbitally Selective Orthogonal Metals. To the extent that these mean field studies reliably capture the ground state in some parameter regime of the microscopic models investigated they 
provide evidence for the occurence of such novel non-Fermi liquid phases in these models. 

\section{Conclusion}
%
We have shown that the slave spin representation provides a natural description of an Orthogonal Metal - a phase that is indistinguishable from the Fermi liquid in conductivity and thermodynamics, but has a sharply different spectral function. The phase transition from a Fermi liquid metal to an Orthogonal Metal proceeds via a critical Fermi surface. In contrast to more traditional slave boson or slave rotor representations \cite{Florens, Lee}, where the death of the Fermi surface is associated with development of a Mott insulator, the slave spin ordering transition does not describe a Mott transition. This is because the ordering transition involves an object (the slave spin) which does not carry the electric charge (or any other quantum numbers of the electron). 

We have shown that the electronic spectral function shows a hard gap in the Orthogonal Metal phase, even though this is a compressible and conducting state of matter. As a result, the transition between the Orthogonal Metal and the Fermi liquid is marked by a dramatic change in the electron spectral function. {We have demonstrated that a continuous transition is possible} in a model with continuous rotation symmetry when Coulomb interactions are included, or in a generic lattice model with $Z_4$ slave spins with either short ranged or Coulomb electron interactions. When the transition is continuous, then when the critical point is approached from the Fermi liquid side, it is marked by a vanishing of the quasiparticle residue. When the critical point is approached from the paramagnetic phase of the slave spins, it is marked by a closing of the electron spectral gap at a critical Fermi surface. We have also provided {two exactly soluble models that realize the Orthogonal Metal phase, which can be accessed from the Fermi liquid via a continuous transition. A critical Fermi surface appears at the phase transition in both models. In the first model, the critical Fermi surface volume obeys the Luttinger count, while in the second model two `mirror' critical Fermi surfaces are present, whose total volume violates the Luttinger count by a factor of two. The emergence of such `mirror' critical surfaces can be traced to a global $Z_2$ symmetry of the second model.} 

We have argued that the prototypical wave-function of the Orthogonal Metal takes the form of a Slater determinant, multiplied by the wavefunction of a paired boson superfluid. This is in contrast to the Slater-Jastrow wavefunction of the Fermi liquid. The form of the Orthogonal Metal wavefunction suggests that an Orthogonal Metal phase may be realized if there is strong pair hopping co-existing with strong electron-electron repulsion. However, the definite identification of an Orthogonal Metal phase in an existing material or `realistic' Hamiltonian remains an open problem. 
 \acknowledgements{We acknowledge useful conversations with Maksym Serbyn, Andrew C. Potter, David F. Mross, Andreas Ruegg and Matthew P. A. Fisher. TS was supported by NSF Grant DMR-1005434. MM was supported by the NSF under Grant No. NSF PHY11-25915.
}

\section{Appendix}
\subsection{$T = 0$ Electrical conductivity in paramagnetic phase of slave spins}
We consider a slave spin ordering transition of the form discussed in \cite{Georges, Si, deMedici, Ruegg}, and we demonstrate that the paramagnetic phase of the slave spins has a non-vanishing conductivity at zero temperature, and thus cannot be interpreted as a Mott insulator. 

We begin with the mean-field treatment of the Hubbard model in the slave-spin representation reviewed in section \ref{sec:slaveintro}. Here one writes the Hamiltonian in terms of fermion and slave-spin operators using Eq.~\ref{eq: repn}. The $f$-fermions and the slave spins are then decoupled in a saddle point approximation and the constraint Eq.~\ref{eq: constraint1} is implemented on average by using a site-independent Lagrange multiplier $\lambda_i = \lambda$. This gives the coupled Hamiltonians \ref{eq: Hf}, \ref{eq: Hi} with renormalized parameters \ref{eq: teff}. 

The slave-spin Hamiltonian \ref{eq: Hi} admits the following phases. When $J_{ij} \gg \lambda$, the slave spins are ferromagnetically ordered, with a non-zero expectation value $\langle \tau^x_i\rangle$. Meanwhile, when $\lambda \gg  J_{ij}$, the slave spins are in a paramagnetic phase, with $\langle \tau^x_i \rangle = 0$. We assume that the $f$ fermions are in a metallic phase, and study the ordering transition in the slave spin sector. How now should we interpret the slave spin ordering transition?

Consider the effective Hamiltonian for the $f$-particles (the charge carriers) in Eq.~\ref{eq: Hf}. For simplicity, let us take the original electron hopping $t_{ij}$ to be only between nearest neighbour sites. 
%
%
The effective $f$-fermion hopping $t'_{ij}$ is given by Eq.\ref{eq: teff} and is non-vanishing provided the nearest neighbour slave spin correlations are non-vanishing, i.e. $\langle \tau^x_i \tau^x_j\rangle \neq 0$. In a single site mean field approximation $\langle \tau^x_i \tau^x_j \rangle \approx \langle \tau^x_i \rangle^2$, the effective hopping does indeed vanish in the paramagnetic phase of the slave spins where $\langle \tau^x_i\rangle = 0$. 
However, once effects beyond the single site mean field approximation are taken into account, the nearest neighbour correlator is non-zero, for any finite values of $U$ and $\lambda$. It follows therefore that once fluctuations beyond mean field are taken into account, the paramagnetic phase of the slave spins has a non-zero electrical conductivity, and should be interpreted as a metallic phase.
Comparison with the Hubbard model Eq.\ref{eq: H} at $V_{ij}=0$ indicates that the $f$-fermion band-structure is identical to the non-interacting tight binding band-structure of the original material, with the bandwidth renormalised by $ \langle \tau^x_i \tau^x_j\rangle$.   The filling factor for $f$ fermions is also the same as the filling factor for the original electron model.

The conductivity may be estimated in the Drude model, and takes the form 
\begin{equation}
\sigma(\omega) = \frac{\nu_0 e^2 v_f^2 \tau}{1 - i \omega \tau}
\end{equation}
where $\nu_0$ is the density of states at the Fermi surface, $v_f$ is the Fermi velocity, and we have introduced a transport time $\tau$. To avoid worrying about the details of the disorder, which govern the transport time, it is convenient to characterize the conductivity by the Drude weight, denoted $\mathcal{D}$, which is defined as the integral of the real part of the conductivity over frequency. The Drude weight is independent of the transport time, and takes the form
\begin{equation}
\mathcal{D} = \pi \nu_0 e^2 v_f^2
\end{equation}
Now we can straightforwardly compare the Drude weights of the non-interacting electron problem, and the interacting problem, decoupled via the slave spin approach. The Fermi velocity has been renormalised by a factor of $ \langle \tau^x_i \tau^x_j\rangle$, and the Fermi surface density of states has been renormalised by $1/(\langle \tau^x_i \tau^x_j\rangle)$. Therefore, if $\mathcal{D}_0$ is the Drude weight of the non-interacting electron probem, then the Drude weight of the interacting problem, decoupled using the slave spin approach, is 
\begin{equation}
\mathcal{D} = \mathcal{D}_0 \times \langle \tau^x_i \tau^x_j\rangle \label{eq: Drude}
\end{equation}
Fermi liquid effects in the $f$-fermion sector may further renormalize $\mathcal{D}$ (see below).

The correlator $\langle \tau^x_i \tau^x_j\rangle$ must be calculated from the slave spin Hamiltonian Eq.~\ref{eq: Hi}. We are interested in the value of the correlator at the slave spin ordering transition, in order to answer the question: is the conductivity significantly suppressed with respect to the non-interacting value when the slave spins disorder?


%

We first consider the case of a one-dimensional Hubbard model. In this case, the zero temperature transverse field Ising model has an exact solution (the Onsager solution). The correlation function may be extracted from this exact solution \cite{Hwang}. We find that the correlation function changes continuously across the critical point, taking value $\langle \tau^x_i \tau^x_j\rangle = 1/\sqrt{2}$ at the critical point of the Ising model. Thus the Drude weight of the one dimensional Hubbard model changes continuously across the ordering transition of the slave spins, and is renormalised relative to the free fermion case by $1/\sqrt2$ at the critical point. Clearly, this should be interpreted as a transition between two distinct metallic phases. 

Meanwhile, for the opposite limit of a three dimensional Hubbard model on a cubic lattice at zero temperature, Ref.\onlinecite{Ruegg} calculated the correlator in a gaussian fluctuations approximation, and found that the correlator changes smoothly across the slave spin ordering transition, taking the value $ \langle \tau^x_i \tau^x_j\rangle \approx 0.1$ at the phase transition. Similar numbers were obtained through cluster dynamical mean field theory calculations \cite{deMedici}. Since three spatial dimensions represents the highest experimentally relevant dimensionality, it follows that the suppression factor $1/ \langle \tau^x_i \tau^x_j\rangle \approx 10$ should be taken as the upper limit for the supression of the Drude weight at the slave spin ordering transition. 

Thus, at the ordering transition of the slave spins, the electrical conductivity is renormalised by a factor $0.1<\langle \tau^x_i \tau^x_j\rangle<0.7$ relative to free fermions. Since this factor is not very small, and since the conductivity changes continuously across the transition, it follows that the ordering transition of the slave spins cannot be interpreted as a metal-insulator transition. It represents a transition between two distinct metallic phases, in one of which the charge carriers are orthogonal to the electrons.

We would like to point out that although our calculation above has been performed in a mean-field approximation, all the conclusions remain qualitatively valid beyond mean-field in the Orthogonal Metal phase. As noted in section \ref{sec:inter}, the Orthogonal Metal is described at low energies as a Landau Fermi-liquid of $f$-fermions. So, as in a Fermi-liquid, the Drude weight may be expressed in terms of the renormalized Fermi-velocity $v_f$ and the forward scattering amplitude $F$ of $f$-fermions. The expression is particularly simple for a rotationally invariant system,
\be {\cal D} = \pi \nu_0 e^2 v^2_f (1 + F_1) \ee
where $F_1$ is the forward-scattering amplitude in the $\ell = 1$ channel. The mean-field calculation above provides an estimate of renormalization of $v_f$ and neglects $F_1$. The latter is expected to be a finite number of $O(1)$ in the vicinity of Fermi-liquid to Orthogonal Metal transition. Thus, the Orthogonal Metal has a finite Drude weight. 

\subsection{Further notes on the transition}
\label{app:PH}
{We would like to point out the difference between the FL-OM transition with $Z_4$ slave spins and certain transitions out of Mott-insulating states, such as, for instance, the well-understood superfluid to Mott-insulator transition in the Bose-Hubbard model.\cite{BoseHubb} 
When driven by a generic perturbation this Mott transition is described by a theory with dynamical exponent $z = 2$. However, when the transition takes place at a constant density it belongs to a different universality class, namely to the $XY$ class with dynamical exponent $z = 1$. Such a priveleged role of the constant density transition is possible because the ground state of the Mott insulating phase carries a fixed density commensurate with the lattice. In contrast, for the Orthogonal Metal to Fermi-liquid transition discussed here, both phases are compressible and commensuration effects are irrelevant. Thus, we expect the transition at fixed (possibly commensurate) density to be described by the same theory in Eq.~\ref{LZ4} as the generic transition. Indeed, as we discussed, the electron number is carried by the $f$-fermions. Therefore, if we neglect the coupling between the slave-spin and the $f$-fermion sectors, the electron density actually stays constant across the slave-spin ordering transition. Once the coupling is present, as we tune the parameter $t$ of the slave-spin theory (\ref{LZ4}) across the transition, the electron density will generically change. However, if one wishes to study a transition at constant density, this change can be compensated by simultaneously tuning the chemical potential of the $f$-fermions.}


\begin{figure}[t]
\includegraphics{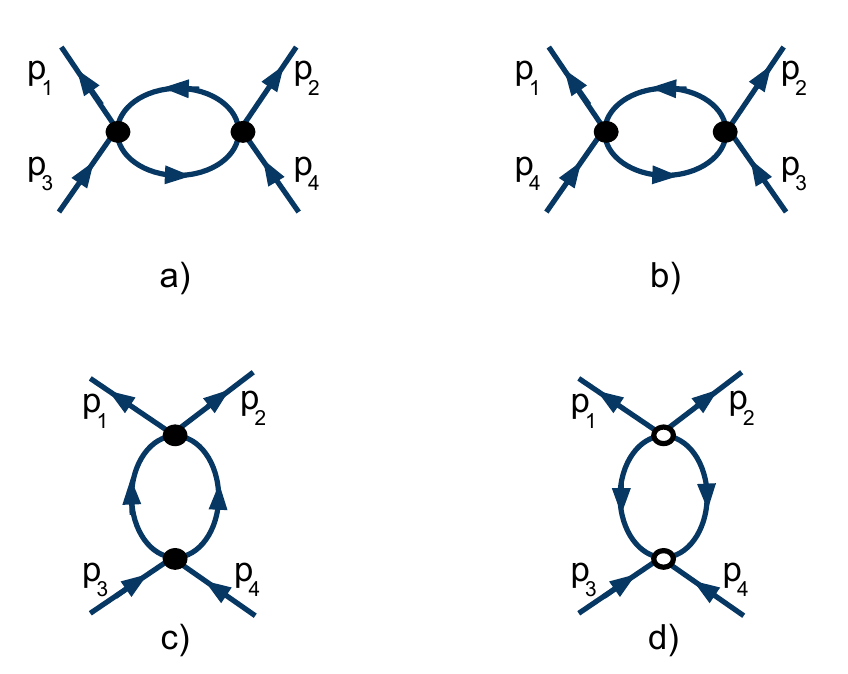}
\caption{One-loop contributions to the four-point function $\Gamma^4(p_1, p_2; p_3, p_4) = \langle \Phi(p_1) \Phi(p_2) \Phi^*(p_3) \Phi^*(p_4)\rangle$ determining the flow of $u$ in the theory (\ref{LZ4}). A solid dot denotes a $u$-vertex and a hollow dot - a $v$-vertex.\label{fig:Feynmu}}
\end{figure}
\begin{figure}[t]
\includegraphics{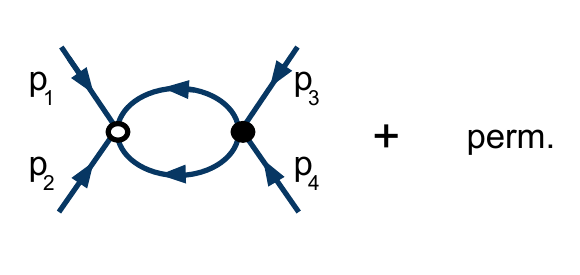}
\caption{One-loop contributions to the four-point function $\bar{\Gamma}^4(p_1, p_2, p_3, p_4) = \langle \Phi^*(p_1) \Phi^*(p_2) \Phi^*(p_3) \Phi^*(p_4)\rangle$ determining the flow of $v$ in the theory (\ref{LZ4}). There are a total of $6$ distinct diagrams obtained
by permutting the external momenta $p_1, p_2, p_3, p_4$.\label{fig:Feynmv}}
\end{figure}
\subsection{RG flow in the ${Z}_4$ slave-spin theory}
\label{app:Z4RG}
In this appendix we discuss the RG flow of the theory (\ref{LZ4}), which describes the ${Z}_4$ slave-spin version of the Orthogonal Metal to Fermi-liquid transition. 

We will focus on spatial dimension $d = 2$, which is the upper critical dimension of the theory (\ref{LZ4}). Here, the coupling constants $u$ and $v$ are marginal at tree-level. Below, we investigate what happens to the flow of these couplings at one-loop level.  Note that when the ${Z}_4$ anisotropy $v$ is set to zero, the theory (\ref{LZ4}) describes the well-understood Bose-Einstein condensation transition.\cite{Sachdev} In this case, if one starts with $u > 0$ then $u$ flows logarithmically to zero, so that the transition is still described by mean-field theory, up to logarithmically suppressed corrections. On the other hand, if one starts with $u < 0$ then $u$ runs away to $-\infty$. This can be interpreted as an instability towards formation of ``molecules" of two $\Phi$ ``atoms". The molecules then Bose-condense before the the individual $\Phi$ atoms condense, i.e. the transition from vacuum to an atomic superfluid proceeds via an intermediate molecular (paired) superfluid phase, which is characterized by $\langle \Phi^2 \rangle \neq 0$ and $\langle \Phi \rangle = 0$. 

We now study how the RG flow described above is modified by the presence of the ${Z}_4$ anisotropy $v$. Note that by applying a phase rotation of $\Phi$, one can choose $v$ to be real and positive. It is convenient to define the following dimensionless coupling constants:
\be \tilde{u} = \frac{m_\Phi u}{4 \pi}, \quad \tilde{v} = \frac{m_\Phi v}{4\pi} \label{eq:dimuv}\ee
The factors of $4 \pi$ are introduced for future convenience.  In addition to the flow of dimensionless couplings (\ref{eq:dimuv}), we expect the field-strength of $\Phi$ and the ``mass" $m_\Phi$ to flow. The latter will lead to a renormalization of the dynamical critical exponent $z$, which once $v \neq 0$ is no longer protected by Gallilean-like invariance. However, at one-loop level both the $\Phi$ field strength and $m_\Phi$ remain unrenormalized, so we focus just on the flow of the couplings (\ref{eq:dimuv}). For briefness, we will drop the tildes on $u$ and $v$ below.

The flow equations for $u$ and $v$ can be obtained from the one-loop corrections to the four point functions $\Gamma^4(p_1, p_2; p_3, p_4) = \langle \Phi(p_1) \Phi(p_2) \Phi^*(p_3) \Phi^*(p_4)\rangle$ and  $\bar{\Gamma}^4(p_1,p_2, p_3, p_4) = \langle \Phi^*(p_1) \Phi^*(p_2) \Phi^*(p_3) \Phi^*(p_4)\rangle$ respectively. The Feynman diagrams for these are displayed in Figs.~\ref{fig:Feynmu},\ref{fig:Feynmv}. Note that the diagrams for $\Gamma^4$ in Figs.~\ref{fig:Feynmu} a),b) vanish for kinematic reasons. Computing the remaining diagrams, we obtain the RG equations,
\bea \frac{d u}{d \ell} &=& - (u^2 + v^2)\label{eq:uFlow}\\
\frac{d v}{d \ell} &=& - 6 u v\label{eq:vFlow}\eea
\begin{center}
\begin{figure}[t!]
\includegraphics[width = 225pt]{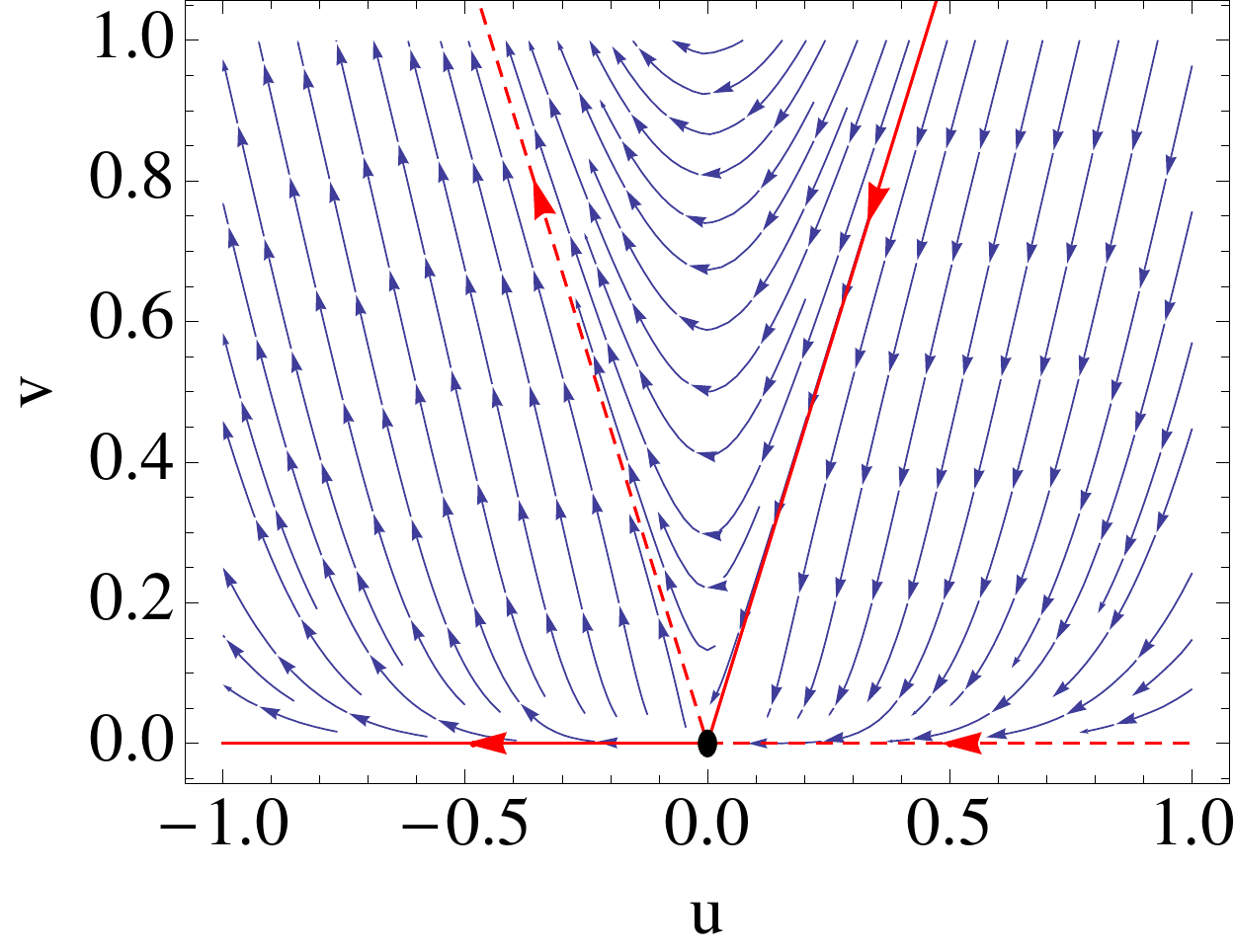}
\caption{RG flow of coupling constants $u$ and $v$ in the theory (\ref{LZ4}). Solid and dashed red lines mark separatrices and attractor lines, respectively.
 \label{fig:Flow}}
\end{figure}
\end{center}
Here $\ell$ parametrizes the rescaling of frequency and momentum, $\omega \to e^{-2 \ell} \omega$, $\vec{q} \to e^{-\ell} \vec{q}$. Fig.~\ref{fig:Flow} displays the RG flow generated by Eqs.~\ref{eq:uFlow},\ref{eq:vFlow}. We note the attractor lines at $u > 0, v = 0$ and $u < 0, v/u = -\sqrt{5}$, as well as the separatrices at $u > 0, v/u = \sqrt{5}$ and $u < 0, v = 0$. We divide the upper $(u,v)$ plane into two regions: region I, defined by $u > 0, v/u < \sqrt{5}$, and its complement. We observe that if the initial couplings lie in region I, the RG flow tends to the attractor line $u > 0, v = 0$, which runs to the Gaussian fixed point $u = 0, v = 0$. The end-stage of this flow can be described analytically: here $v \ll u$, so we can drop the ${v}^2$ term in Eq.~\ref{eq:uFlow}. Then,
\be u(\ell) = \frac{u_0}{1 + u_0 \ell}, \quad  v(\ell) = \frac{v_0}{(1 + u_0 \ell)^6}\ee
Thus, both $u$ and $v$, indeed, flow logarithmically to zero. Note that for $\ell \to \infty$, $u \sim \ell^{-1}$ and $v \sim \ell^{-6}$. Thus, $v \ll u$ and our assumption is justified.

Thus, if the initial couplings lie in region I, the transition will be described by mean-field theory up to logarithmically suppressed corrections. On the other hand, if the initial couplings lie outside of region I, the RG flow tends towards the attractor line $u < 0, v/u = -\sqrt{5}$, which runs away to $u = -\infty, v = - \infty$. By analogy with the physical interpretation of the $u \to -\infty$ RG flow in the theory with $v = 0$, we may hypothesize that the present runaway flow signals formation of pairs of $\Phi$-particles. The ordered phase is then characterized by a pair-condensate, $\langle \Phi^2 \rangle \neq 0$, while having $\langle \Phi \rangle = 0$. When discussed in the context of the original electronic theory, this phase is nothing but a ${Z}_2$ slave-spin Orthogonal Metal. Thus, in this scenario, for initial couplings outside of region I, the transition from a ${Z}_4$ slave-spin Orthogonal Metal to a Fermi-liquid occurs via an intermediate ${Z}_2$ slave-spin Orthogonal Metal phase.


\begin{thebibliography}{99}
\vspace{-7mm}

\providecommand{\natexlab}[1]{#1}
\providecommand{\url}[1]{\texttt{#1}}
\expandafter\ifx\csname urlstyle\endcsname\relax
  \providecommand{\doi}[1]{doi: #1}\else
  \providecommand{\doi}{doi: \begingroup \urlstyle{rm}\Url}\fi
  
\bibitem[Senthil (2008)]{Senthil}
T. Senthil, Phys. Rev. B 78, 035103 (2008) 

\bibitem[Georges (2005)]{Georges}
L. de'Medici, A. Georges and S. Biermann, Phys. Rev. B 72, 205124 (2005)

\bibitem[Si (2010)]{Si}
Rong Yu and Qimiao Si, Phys. Rev. B 84, 235115 (2011)

\bibitem[deMedici (2010)]{deMedici}
S. R. Hassan and L. deÕ Medici, Phys. Rev. B 81, 035106 (2010) 

\bibitem[Ruegg (2010)]{Ruegg}
A. Ruegg, S.D.Huber and M. Sigrist, Phys. Rev. B 81, 155118 (2010)

\bibitem[Fabrizio (2005)]{Fabrizio}
M. Schiro and M. Fabrizio, Phys. Rev. B 83, 165105 (2011)

\bibitem[Florens (2004)]{Florens}
S. Florens and A. Georges, Phys. Rev. B 70, 035114 (2004)

\bibitem[Lee (2006)]{Lee}
P.A. Lee, N. Nagaosa and X-G. Wen, Rev. Mod. Phys. 78, 17Ð85 (2006)

\bibitem{LGWstar} A. V. Chubukov, T. Senthil, and S. Sachdev, Phys. Rev. Lett. 72, 2089 (1994);
S. V. Isakov, T. Senthil, and Y. B. Kim, Phys. Rev. B 72, 174417 (2005);
T. Senthil and O. Motrunich,
Phys. Rev. B 66, 205104 (2002);
 T. Grover and T. Senthil, Phys. Rev. B 81, 205102 (2010);
S. V. Isakov, R. G. Melko, and M. B. Hastings, Science 335, 193 (2012). 



\bibitem[Kondo1 (2003)]{Kondo1}
T. Senthil, S. Sachdev and M. Vojta, Phys. Rev. Lett. 90, 216403 (2003)

\bibitem[Kondo2 (2006)]{Kondo2}
T. Senthil, M. Vojta and S. Sachdev, Phys. Rev. B, 69, 035111 (2004)

\bibitem{paul}I. Paul, C. Pepin, and M. R. Norman, 
Phys. Rev. Lett. 98, 026402 (2007) . 

\bibitem[Mott (1998)]{Mott}
T. Senthil, Phys. Rev. B 78, 045109 (2008) 

\bibitem{Mott3d}Daniel Podolsky, Arun Paramekanti, Yong Baek Kim, and T. Senthil,
Phys. Rev. Lett. 102, 186401 (2009).

\bibitem[Ruegg2 (2011)]{Ruegg2}
A. Ruegg and G.A.Fiete, Phys. Rev. Lett. 108, 046401 (2012)

\bibitem[Kane (2005)]{Kane}
C.L.Kane and E. J. Mele, Phys. Rev. Lett. 95, 226801 (2005)

\bibitem[Hwang (1963)]{Hwang}
Kerson Huang, \emph{Statistical Mechanics}, Wiley (1963)



\bibitem[TSenthil (2000)]{TSenthil}
T. Senthil and M.P.A. Fisher, Phys. Rev. B 62, 7850-7881 (2000)

\bibitem[Motrunich (2005)]{Motrunich}
O.I. Motrunich and T. Senthil, Phys. Rev. B 71, 125102 (2005)

\bibitem[Altland (2006)]{Altland}
A. Altland and B.D. Simons, \emph{Condensed Matter Field Theory}, Chap. 8, Cambridge University Press (2006)

\bibitem[Mahan (1990)]{Mahan}
Gerald D.Mahan, \emph{Many Particle Physics, $2^{nd} ed$}, Chap. 7. Plenum Press (1990)

\bibitem[Giamarchi (2005)]{Giamarchi}
T. Giamarchi, {\it Quantum Physics in One Dimension}, Clarendon Press (2005)

\bibitem[Sachdev (1999)]{Sachdev}
S. Sachdev, \emph{Quantum Phase Transitions}, Cambridge University Press (1999)


\bibitem[Morinari (2002)]{Morinari}
S. Sachdev and T. Morinari, Phys. Rev. B 66, 235117 (2002)

\bibitem{maissam} A more sophisticated example is a continuous transition between superfluid and quantum Hall states of bosons on a lattice that is studied in 
M. Barkeshli and J. McGreevy, http://arxiv.org/abs/1201.4393. 

\bibitem{BoseHubb}
M.P.A. Fisher, P.B.Weichman, G.Grinstein, D.S. Fisher, Phys. Rev. B {\bf 40}, 546 (1989) 

\bibitem[Casey (2012)]{Casey}
P.A. Casey and P.W. Anderson, Phys Rev Lett, 106, 097002 (2011)

\bibitem[Shastry (2012)]{Shastry}
 B.S. Shastry, arxiv:1110.1032v2 (2012)

\bibitem[Wu (1982)]{Wu}
F.Y.Wu, Rev. Mod. Phys. 54, 235Ð268 (1982)

\bibitem[Assad (1997)]{Assad}
F.F.Assaad, M. Imada and D. J.Scalapino, Phys. Rev. B 56, 15001 (1997)

\end{thebibliography}
\end{document}